\documentclass[a4paper, 12pt, british, twoside]{article}

\usepackage[left=2.5cm, right=2.0cm, top=2.5cm, bottom=2.0cm, ]{geometry}

\usepackage[utf8]{inputenc}
\usepackage[T1]{fontenc}
\usepackage[width=0.9\textwidth]{caption} 
\usepackage[ngerman, main=british]{babel}
\usepackage[autostyle]{csquotes}

\usepackage{amsmath, amssymb, amsfonts, mathtools}
\usepackage{booktabs, lmodern}

\usepackage{tikz,pgfplots}
\pgfplotsset{compat=newest}
\graphicspath{./pics}
\definecolor{hzbblue}{cmyk}{1.0,0.50,0,0.18}
\definecolor{hzbcyan}{cmyk}{1.0,0.0,0.0,0.2}
\definecolor{hzbgreen}{cmyk}{0.35,0.0,1.0,0.0}
\definecolor{darkblue}{cmyk}{1.0, 0.5, 0, 0.35}
\definecolor{darkgreen}{rgb}{0.0, 0.5, 0.0}
\newcommand{\Rb}{\ensuremath{\mathit{Rb}}}
\newcommand{\Ba}{\ensuremath{\mathit{Ba^{\!+\!}}}}

\usepackage[backend=biber, style=phys, doi=true, biblabel=brackets,natbib=true]{biblatex}
\usepackage[hyperindex, bookmarks, pdfdisplaydoctitle,
]{hyperref}
 
\providecommand{\braket}[2]{{\ensuremath{\left< { #1 } {\mid} { #2 } \right>}}}
\providecommand{\ketbra}[2]{ {\ket{ #1 }\!\bra{ #1 }} }
\providecommand{\op}[1]{\ketbra{#1}{#1}}

\providecommand{\1}{\mathbb{1}}
\providecommand{\ket}[1]{ \ensuremath{\left| #1 \right>} }
\providecommand{\bra}[1]{ \ensuremath{\left< #1 \right|} }

\providecommand\bohr[1][]{ {\ensuremath{a_{\!B}^{#1}}} }
\providecommand\hartree[1][]{ {\ensuremath{E_{\!H}^{#1}}} }\providecommand\fs{ {\ensuremath{\mathrm{fs}}} }\providecommand\eV{ {\ensuremath\mathrm{eV}} }

\providecommand\distance{{\mathcal{R}}}
\providecommand{\vol}{{\mathcal V}}

\providecommand{\txtrev}[1]{
	{\begingroup#1\endgroup}%
}

\begin{document}

\renewcommand\sectionautorefname{Section}
\let\subsectionautorefname\sectionautorefname
\let\subsubsectionautorefname\sectionautorefname

\date{revised 19\textsuperscript{th} April 2024}
\title{Time-Resolved Rubidium-Assisted Electron Capture by Barium (II) Cation}
\author{
\begin{minipage}{0.95\textwidth}
    \renewcommand{\theenumi}{\textsuperscript{\alph{enumi}}}
    \renewcommand\labelenumi\theenumi
    \normalsize
	Axel Molle,\ref*{kul}$^,$\ref*{csm}$^,$\ref*{hzb}$^,$\ref*{fub}$^,$\thanks{\href{mailto:axel.molle@kuleuven.be}{axel.molle@kuleuven.be}} 
	Jan Philipp Drennhaus,\ref*{kul} 
	Viktoria Noel,\ref*{hdp}$^,$\ref*{ox}$^,$\ref*{hzb}
	Nikola Kolev,\ref*{ucl}$^,$\ref*{edi}$^,$\ref*{hzb}\linebreak
	and Annika Bande\ref*{hzb}$^,$\ref*{luh}$^,$\thanks{annika.bande@helmholtz-berlin.de}
\footnotesize\smallskip
	\begin{enumerate}
        \addtolength\itemsep{-.25em}
\item 
		Department of Physics and Astronomy, KU Leuven
		Celestijnenlaan 200d, 
		3001 Leuven, Belgium%
		\label{kul}
        \item Department of Physics, Colorado School of Mines, 1301 19\textsuperscript{th} St., Golden, CO 80401, USA
        \label{csm}
		\item Theory of Electron Dynamics and Spectroscopy, Helmholtz-Zentrum Berlin für Materialien und Energie, 14109 Berlin, Germany \label{hzb}
		\item Institute of Chemistry and Biochemistry, Freie Universit\"at Berlin, 14195 Berlin, Germany \label{fub}
        \item Institute for Theoretical Physics, Heidelberg University, 69120 Heidelberg, Germany \label{hdp}
		\item St Edmund Hall, University of Oxford, Oxford OX1 4AR, United Kingdom \label{ox}
		\item London Centre for Nanotechnology and Department of Electronic and Electrical Engineering, University College London, London WC1H 0AH, United Kingdom \label{ucl}
		\item University of Edinburgh, School of Physics and Astronomy, Edinburgh EH9 3FD, U.K. \label{edi}
        \item Institute of Inorganic Chemistry, Leibniz University Hannover, Callinstr. 9, 30167 Hannover, Germany \label{luh}
	\end{enumerate}
\end{minipage}
}

\maketitle
\vspace*{-2.7em}

 	\begin{abstract}\footnotesize
Non-local energy transfer between bound electronic states close to the ionisation threshold is employed for efficient state preparation in dilute atom systems from technological foundations to quantum computing.
The generalisation to electronic transitions into and out of the continuum is lacking \txtrev{quantum} simulations necessary to motivate such \txtrev{potential} experiments.
    Here, we present the first development of an electron-dynamical model simulating fully three-dimensional atomic systems for this purpose. We investigate the viability of this model in its current stage for the prototypical case of recombination of ultracold barium~(II) by environment-assisted electron capture thanks to a rubidium atom in its vicinity. Both atomic sites are modelled as effective one-electron systems using the Multi Configuration Time Dependent Hartree (\textsc{mctdh}) algorithm and can transfer energy by the dipole-dipole interaction.

  We find that the simulations are robust enough to realise assisted electron capture over dilute interatomic distance through excitation and ionisation of rubidium which we are able to quantify by comparing simulations with and without interatomic energy exchange. 
  For our current parameters, 
not yet optimised for reaction likelihood,
  an environment-ionising assisted capture has a probability of $1.9\times10^{-5}~\%$ over the first $15~\mathrm{fs}$ of the simulation. The environment-exciting assisted-capture path to $[\text{Ba}^{+*}\text{Rb}^{*}]$ appears as a stable long-lived intermediate state with a probability of $8.2\times10^{-4}~\%$ for at least $20~\mathrm{fs}$ after the capture has been completed. 

  \txtrev{This model shows potential to predict optimised parameters as well as to accommodate the conditions present in experimental systems as closely as possible. We put the presented setup forward as a suitable first step to experimentally realise environment-assisted electron capture with current existing technologies.}
	\end{abstract}

\section{Introduction}

\begin{figure}[h]
    \centering
    \includegraphics{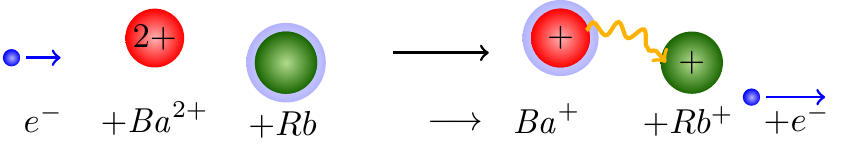}
    \caption{Schematic representation of interatomic Coulombic electron capture (\textsc{icec}) for a hetero-atomic system. In this scenario, an incoming electron $e^-$ is captured by a barium~(II) ion Ba$^{2+}$, assisted by a nearby rubidium atom Rb. Simultaneously, the transferred excess energy ionises the Rb. Various intermediate reaction channels can lead to the same final state which is discussed in \autoref{s:channels} in more detail.}
    \label{f:icec}
\end{figure}

When free electrons attach to atoms, ions and molecules, they usually lose their excess energy by intra-atomic or intramolecular channels such as the emission of photons, the instigation of vibrations and rotations or the excitation of electrons within the encountered system. Environment-assisted electron capture was proposed as a competing channel in 2009,\autocite{gokhberg-jphysb-2009} when it was shown that an incoming electron can attach to the capturing cation by transferring part of its energy to an atom in the vicinity of the cation. This process was given the name interatomic Coulombic electron capture (\textsc{icec}), and is represented in \autoref{f:icec}.

\paragraph{}
\textsc{icec} has been shown to affect systems ranging from halogen atom/anion pairs,\autocite{gokhberg-jphysb-2009} clusters of noble gas atoms,\autocite{gokhberg-physrev-2010} and nano-electronic devices,\autocite{pont-physrev-2013,bande-epjconf-2015,pont-jphys-2016,molle-jcp-2019,pont2019-prepare} to biologically relevant microhydrated minerals as well as organic $\pi$ systems.\autocite{molle-jcp2023, gokhberg-physrev-2010} It is therefore a fundamental energy transfer process.
Recent findings from accurate time-independent computations show that environment-assisted electron capture is more efficient\autocite{sisourat-physreva-2018,molle-pra2021} than initially estimated\autocite{gokhberg-physrev-2010} and will therefore potentially play a bigger role in low-energy interactions than originally assumed.

Understanding this process can be important for a variety of fields.\autocite{bande-jphysb2023}
For instance, in biological organisms slow electrons can lead to the harmful propagation of irradiation damage through dissociative electron attachment. This breaks chemical bonds, and consequently destroys molecules and DNA.\autocite{boudaiffa-science-2000,sanche-radphyschem-2016,hergenhahn-intjradbio-2012} 
It was theoretically shown that \textsc{icec} is a competing process with other typical electronic processes. This was demonstrated in the biologically relevant ion $\mathrm{Mg}^{2+}$ near water molecules.\autocite{gokhberg-jphysb-2009,gokhberg-physrev-2010} The range of examples has been extended recently to other bio-relevant alkali and alkaline earth cations where it has been found that \textsc{icec} is dominant over much longer distances than initially anticipated such that it might act macroscopically beyond the typical sizes of the distinguishable hydration shells.\autocite{molle-jcp2023} 
It is furthermore predicted to be relevant in solid-state systems   
\autocite{pont-physrev-2013,pont-jphys-2016} as a possible concern in the context of modern electronic devices constructed from semiconductor quantum dots, which emerged from the constant race for more minute components in integrated circuits.\autocite{molle-jcp-2019,pont2019-prepare}
Through these theoretical investigations across different systems, \textsc{icec} has been established as a fundamental energy-transfer process. However, it lacks dedicated experimental investigations to date.\autocite{bande-jphysb2023}

\paragraph{}
Here we propose dilute ultracold atomic systems as a viable candidate for such investigations.
Although theoretical studies on \textsc{icec} have already targeted atoms immersed in helium clusters,\autocite{sisourat-physreva-2018} dilute ultracold atoms have not yet been discussed in this context. Technologies have advanced strongly over the last decades, and ultracold atoms trapped in optical lattices and optical tweezers make up an essential part of today's atomic, molecular and optical physics. They are used for the construction of e.g.~high-precision atomic clocks,\autocite{bothwell-nat2022} ultracold electron sources,\autocite{PhysRevAccelBeams.22.023401} and potential candidate systems for quantum computing processors. 
Given the current available experimental control, we consider ions in ultracold atom clouds as a type of system which can already be approached experimentally regarding the \textsc{icec} process.
\txtrev{Eventually, we wish to motivate the experimental observation of ICEC in the future. Since modern cold atomic experiments are time resolved,\autocite{kruekow-physrevA2016, PhysRevAccelBeams.22.023401,wolf-science2017,sikorsky-natcomm2018,bahrami-physstatsol2019} we adhere to this framework and employ a time-resolved simulation in our numerical study.}
We present therefore the first development of a time-resolved quantum-mechanical simulation of the electron dynamics of environment-assisted electron capture between atoms. We choose the prototypical system of a barium~(II) cation, Ba$^{2+}$, with a nearby rubidium atom, Rb, since comparable systems are already considered experimentally.\autocite{kruekow-physrevA2016, PhysRevAccelBeams.22.023401}

In practice, entrapment of cold atoms and ions has been undertaken for a while.
Hybrid traps employing two different trap techniques for ions and atoms simultaneously are the current state of the art such that interactions of captured ions within ultracold atom clouds are investigated in real time.\autocite{wolf-science2017,sikorsky-natcomm2018,bahrami-physstatsol2019} 
The low temperatures in these experiments allow for the investigation of numerous low energy phenomena such as fundamental Coulomb interactions, time-resolved selective state-to-state chemistry and problems from astrophysics and astrochemistry.

To our knowledge, the three necessary components for realising \textsc{icec}, namely the ion, an electron source and the assisting atom are not yet brought together but are at our fingertips in experimental technologies. The stable entrapment of ultracold atoms builds the foundation of modern ultracold electron sources.\autocite{PhysRevAccelBeams.22.023401} We argue therefore that they are a feasible extension compatible with prevalent hybrid cold ion-atom systems.\autocite{kruekow-physrevA2016,bahrami-physstatsol2019}
We are developing the presented quantum-mechanical model to eventually motivate this experimental addition, despite its challenges.
Hybrid-trap vacuum chambers are sensitive to disturbances. Assembling all experimental components to achieve stable ultra high vacuum ($<10^{-6}$~Pa) at low temperatures is a challenge by itself: it requires multi-stage pumping and ion gauge measurement, and particular procedures like baking of the chamber to remove traces of gases.
Nevertheless, electron sources were discovered about two centuries ago\autocite{faraday-ptrsl1838} and have been commercially used for a long time e.g.~in cathode rays for gas discharge lamps, oscilloscopes, and televisions. They are conceptually simple and thus also simple enough to add into modern experiments with modular design. Moreover, ultracold electron beams are produced by near-threshold optical ionisation of optically trapped ultracold atoms,\autocite{PhysRevAccelBeams.22.023401} and are therefore compatible with modular trapped atom and ion interaction experiments.\autocite{kruekow-physrevA2016} As such, we consider it only a matter of time before the first realisation of an ultracold \textsc{icec} experiment is reported.

\paragraph{}
This paper therefore explores the feasibility of time-resolved electron-dynamical simulations of environment-assisted electron capture in \txtrev{ultracold atoms. 
It proposes a theoretical model for further development that may be explored for a potential experimental observation}. We begin by an outline of expectations regarding the possible reaction channels (\autoref{s:channels}) before presenting the quantum-mechanical constituents of the dynamical model under development.
We will present our model by introducing the underlying coordinate system in \autoref{s:model}, that influences the expression of the Hamiltonian operator in first quantisation and the approach for the time-dependent wavefunction in the \textit{multi-configuration time-dependent Hartree} (\textsc{mctdh}) formalism, \autoref{s:theory-hamiltonian} and \autoref{s:theory-wavefunction}, respectively. We will conclude the theory section with a summary of the initial preparation of the wavefunction as starting point for the time evolution.
The computational details including the chosen physical and technical parameters for the electron dynamics calculation with the Heidelberg \textsc{mctdh} software package are given thereafter in \autoref{s:compdetails}. 
We present the numerical results of the simulation prior to a discussion of their interpretation in \autoref{s:results}. 
We illustrate the evolution of electron probability density with respect to the three continuum coordinates first to investigate the electron attachment to $\mathrm{Ba}^{2+}$ and the electron emission or excitation at $\mathrm{Rb}$ (cf. \autoref{s:results-DensEvol}). Secondly, we investigate the probability flux of the continuum electron passing through the boundary of the simulated interaction volume in temporal resolution in \autoref{s:results-flux}. Lastly, we calculate reaction probabilities for rubidium-assisted electron capture where we distinguish between emission and excitation of the rubidium electron (cf. \autoref{s:reaction-prob}). Our summarising interpretation is discussed (\autoref{s:Discussion}) regarding the impact of these results before we conclude (\autoref{s:Conclusion}) the findings of the presented development of the quantum-dynamical model for assisted electron capture in dilute atomic systems.

Throughout this manuscript, we consistently consider the subsystem of a single outer valence electron of a barium~(II) in dynamic interaction with the assisting subsystem of a single outer valence electron of a distanced rubidium~(I). In our aim to develop a simple intuitive model covering major aspects of the interatomic process, both subsystems will conserve their individual net-charge but be allowed to occur in their electronic ground state, in their electronically singly-excited bound state, or in their ionised form with one associated moving free electron. In the case of the barium~(II) subsystem for instance, we will thus speak of barium~(II) -- i.e.~[Ba$^+$],
of excited barium~(II) -- i.e.~[(Ba$^+$)$^*$],
and of the barium~(II) cation with a free electron -- $[\mathrm{Ba}^{2+} + e^-]$. \txtrev{We wish to point out for clarity that the doubly-ionized barium, Ba$^{2+}$ on its own is indeed a barium~(III) which is being used in conjunction with the dynamically modeled additional outer valence electron that make up this charge-conserved barium~(II) subsystem altogether.}
Describing the dynamics, we might wish to distinguish an observation on the evolution at the one site from a simultaneous development at the interacting other site. For the sake of aided readability through our description, we may therefore simply speak of the \textit{barium} or the \textit{rubidium} site while we still consistently mean the barium~(II) and rubidium~(I) subsystem with their varying electronic states of their respectively associated single outer valence electron.
\txtrev{We would like to emphasise that \textsc{icec} is about electronic excitation transfer with continuum states, and not merely an electron transfer in a stable ground state of barium~(II) and rubidium.}

\section{Theory}\label{s:theory}
Prior investigations on the electron dynamics of \textsc{icec} were dealing with systems with up to three bound levels in artificial atoms known as quantum dots, which were confined in one dominant transport direction by embedding in a nanowire.\autocite{pont2019-prepare,molle-jcp-2019} This work represents a significant step from previous studies on assisted electron attachment in nanowire-embedded quantum dots to a quantum-dynamical simulation of energy transfer between atoms. 
The transition to real atoms will offer an abundance of capturing and assisting electron orbitals. This multitude of accessible states will play a significant difference and it has been shown that it cannot be reduced back to a few-level treatment.\autocite{molle-jcp2023}

In the following subsection, we are therefore presenting which reaction channels are expected and need to be included in a successful quantum-dynamical model which is laid out thereafter.

\subsection{The Reaction Channels}\label{s:channels}
We wish to simulate the standard process of recombination of an electron with a barium~(II) cation Ba$^{2+}$ upon energy transfer to the assisting outer valence electron of a neutral rubidium atom, Rb.  
In vacuum, an incident free electron can only scatter elastically in the form of Rutherford scattering, or inelastically at the barium~(II) cation by radiative emission of energy.
The capture of the electron to the ground state of barium~(II), Ba$^+$, necessitates a simultaneous emission of a photon:
\begin{equation}
e^- + \mathrm{Ba}^{2+}  \longrightarrow \mathrm{Ba}^{+} + \gamma.  \;
\end{equation}
Similarly, this process would allow a capture into excited states, or allow for higher-order radiative processes involving e.g.~the capture in an excited state under emission of a first photon (i.e. initial photorecombination) and subsequent relaxation by emission of a second photon (i.e. secondary photorelaxation).

As we introduce a rubidium atom into the neighbourhood, energy transfer between the recombining cation and the nearby atom allows for additional capture mechanisms. 
This energy transfer is a non-radiative process that has been shown to be dominant in many systems over photorecombination,\autocite{gokhberg-jphysb-2009,mueller-physrevlett-2010,voitkiv-physrev-2010} and is inherently long-ranged.\autocite{molle-jcp2023} 
As this paper deals with assisted electron capture, we will only focus on these non-radiative rubidium-assisted processes. The transferred excess energy from the electron capture could also be absorbed into kinetic energy of the heavy rubidium nucleus which will not be covered in this paper due to the different time scales of electronic and heavy-core wavefunctions stemming from their large differences in magnitude of their respective mass. The atomic nuclei will be treated as stationary and we concentrate on environment-assisted capture by energy transfer between the outer valence electrons of both sites.

\begin{figure}[h]
    \centering
    \includegraphics{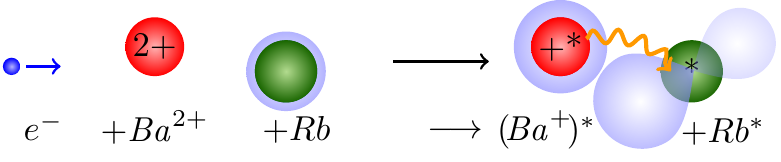}
    \caption{Schematic representation of an intermediate step in assisted electron capture for a hetero-atomic system. In this scenario, an incoming electron $e^-$ is captured by a barium~(II) ion Ba$^{2+}$, assisted by a nearby rubidium atom Rb. Simultaneously, the transferred excess energy excites the Rb atom such that (temporarily) both species exist in an excited state. Depicted by the purple (electron) clouds, the rubidium electron initially in the $5s$ orbital is excited into a $p$ orbital. This excited state can then lead to various further reactions, discussed in more detail in context of Eq.~\eqref{eq:Ba+*Rb*}.}
    \label{f:excited-capture}
\end{figure}

\paragraph{}
The \textsc{icec} process starts with the incident electron of kinetic energy $\epsilon$ getting captured into the ground state or an excited state of the barium~(II), Ba$^+$. 
The ground states of Ba$^+$ and Rb have binding energies of $10.00$ and $4.18$~eV, respectively.\autocite{NIST:ASD,curry-jpcrd2004, johansson-arkfys1961}
In a ground-state capture, $10 \ \mathrm{eV} +\epsilon$ get transferred to the Rb atom which will lead to its ionisation, known as \textit{direct \textsc{icec}}(see \autoref{f:icec}),\autocite{molle-jcp-2019,pont2019-prepare}
\begin{equation}
e^- + \mathrm{Ba}^{2+} + \mathrm{Rb} \longrightarrow \mathrm{Ba}^{+} +\mathrm{Rb}^{+} + e^-.
\label{eq:Ba+Rb+}
\end{equation}
If the electron gets captured into an excited state instead, the transferred energy will be lower. The difference in transferred energy reflects the difference in the binding energy of the capturing excited state with respect to the ground state of barium~(II), 
\begin{equation}
e^- + \mathrm{Ba}^{2+} + \mathrm{Rb} \longrightarrow (\mathrm{Ba}^{+})^* + \mathrm{Rb}^+ +e^-.
\label{eq:Ba+*Rb+}
\end{equation}
If $\epsilon$ is smaller than the binding energy of the Rb ground state of 4.18~eV, this does not only lead to the ionisation of rubidium,
but can also lead to rubidium's excitation for appropriate excited states of barium~(II)
\begin{equation}
e^- + \mathrm{Ba}^{2+} + \mathrm{Rb} \longrightarrow (\mathrm{Ba}^{+})^* + \mathrm{Rb}^*, 
\label{eq:Ba+*Rb*}
\end{equation}
as depicted in \autoref{f:excited-capture}. If the Rb atom then de-excites via emission of a photon, this is known as \textit{two-centre dielectronic recombination} (\textsc{{\footnotesize2}cdr}),\autocite{jacob-pra2019,eckey-pra2018,voitkiv-physrev-2010,mueller-physrevlett-2010} or alternatively as \textit{resonant interatomic Coulombic electron capture} (r\textsc{icec}).\autocite{gokhberg-physrev-2010}

As we focus on non-radiative processes here, we note that the excess energy stored in $\mathrm{Rb}^*$ can leave the system by being transferred back to $(\mathrm{Ba}^{+})^*$ via \textit{interatomic Coulombic decay} (\textsc{icd}) \autocite{cederbaum-physrevlett-1997} where it leads to the reemission of the captured electron, in the sense of a delayed scattering.\autocite{eckey-pra2018} 
Alternatively, the excitation energy of $(\mathrm{Ba}^{+})^*$ can transfer similarly through \textsc{icd} to rubidium leading to a delayed \textsc{icec} with ionisation, previously described as \textit{resonance-enhanced \textsc{icec}}.\autocite{molle-jcp-2019,pont2019-prepare,remme-jphysb2023} 
We note that the final state [Ba$^+$Rb$^+$] represents the overall ground state of the considered doubly charged diatomic system, and is thereby lower in energy than the initial state [Ba$^{2+}$Rb] with a neutral rubidium atom. Their energy difference is mitigated by the energy difference between the incident and the outgoing free electron.
Hence, these interatomic relaxation processes both include a secondary energy transfer. Their decay width could be narrow though depending on their atomic resonance character. The system could therefore, potentially, stay for a relatively long time in this singly-charged diatomic excited state [(Ba$^+$)$^*$Rb$^*$]. 

\paragraph{}
Lastly, the incoming electron could diffract inelastically at Ba$^{2+}$ for appropriate incident energies under transfer of some of its kinetic energy to rubidium. This may lead to the direct ionisation of Rb due to the interatomic energy transfer for large enough incident electron energies,
\begin{equation}\label{eq:Ba2+Rb+}
e^- + \mathrm{Ba}^{2+} + \mathrm{Rb} \longrightarrow e^- + \mathrm{Ba}^{2+} + \mathrm{Rb}^+ +e^- , 
\end{equation}
or it may lead to subsequent excitation of rubidium,
\begin{equation}\label{eq:Ba2+Rb*}
e^- + \mathrm{Ba}^{2+} + \mathrm{Rb} \longrightarrow 
e^- + \mathrm{Ba}^{2+} + \mathrm{Rb}^*.
\end{equation}
The excited rubidium atom can then, in turn, relax radiatively or by transferring its excess energy back to the scattered electron with a certain interatomic decay rate.
Numerically, we will have to find a compromise in an incident electron wave packet that is fast enough to be passing through the simulation in a reasonable time span, yet slow enough to prohibit ionising rubidium through inelastic diffraction according to Eq.~\eqref{eq:Ba2+Rb+}.
We will avoid the ionisation of rubidium by keeping the incident electron energy the significant majority of the incident wavepacket below 4.18~eV, the ionisation threshold of Rb.\autocite{NIST:ASD,johansson-arkfys1961,lorenzen-physscr1983} The first electronic excitation of rubidium lies 1.56~eV above its ground state.\autocite{sansonetti-jpcrd2006} We will not be able to avoid exciting rubidium through inelastic diffraction according to Eq.~\eqref{eq:Ba2+Rb*} in this work.

\subsection{The Quantum Dynamical Model}\label{s:model}
\begin{figure}[!htb]
    \centering
	\includegraphics[width=0.5\textwidth]{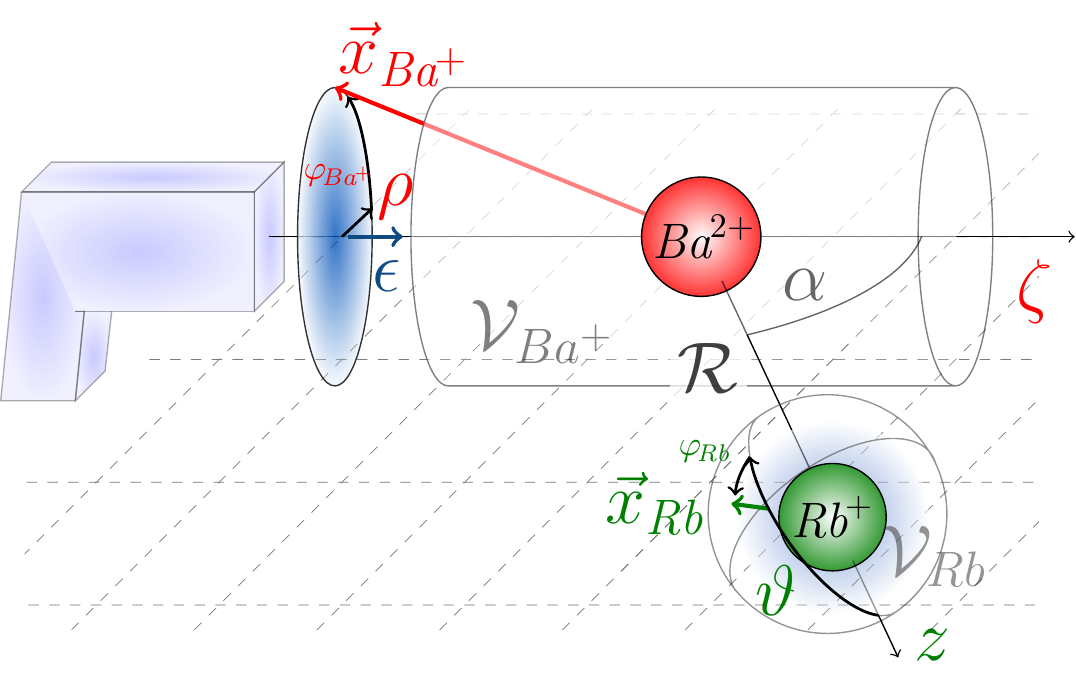}
	\caption{Schematic coordinates attached to an interatomic Coulombic electron capture by a barium~(II) cation, Ba$^{2+}$ in proximity to a neutral rubidium atom at relative position $\vec{\distance}$ from the cation. The incoming electron with kinetic energy $\epsilon$ has position $\vec{x}_\Ba$ relative to the Ba$^{2+}$ ion core given by the longitudinal distance $\zeta$, transverse distance $\rho$ and azimuthal angle $\varphi_\Ba$. The position $\vec{x}_\Rb$ of the assisting electron bound locally to Rb is given by the spherical polar coordinates of radial distance $r$ to the atomic core, the longitudinal angle $\vartheta$ and azimuthal angle $\varphi_\Rb$.}
    \label{f:coordinates}
\end{figure}
Consider a barium~(II) cation, Ba$^{2+}$  and a neutral rubidium atom at relative position $\vec{\distance}$ as indicated in \autoref{f:coordinates}. The cation interacts with an incident electron whose position relative to the barium ion core, 
\begin{equation}
    \vec{x}_\Ba = (\zeta, \rho, \varphi_\Ba)
    \label{eq:xBa}
\end{equation}
is given in \textit{cylindrical coordinates} by its longitudinal distance $\zeta$, its transverse distance $\rho$, and its azimuthal angle $\varphi_\Ba$. The incident electron initially moves parallel to the $\zeta$ axis.

The position of the outer valence electron of rubidium,
\begin{equation}
    \vec{x}_\Rb = (r,\vartheta,\varphi_\Rb)
    \label{eq:xRb}
\end{equation} can be expressed in \textit{spherical polar coordinates} by its radial distance $r$ from the rubidium core, its longitudinal angle $\vartheta$ from the $z$ axis and its azimuthal angle $\varphi_\Rb$. The boundaries of the volumes $\vol_\Ba$ and $\vol_\Rb$, which will allow for the observation of the scattered electron on Ba$^{2+}$ and of the emitted electron on Rb respectively, are also depicted schematically in \autoref{f:coordinates}. 

\subsubsection{The Hamiltonian}\label{s:theory-hamiltonian}
We model the ion-atom system as two effective one-electron systems that \txtrev{exchange energy via the long-range interaction $\hat V_{12}$}. The Hamiltonian of the total system is given by
\begin{equation}
	\hat{H} := \hat{h}_\Ba + \hat{h}_\Rb + \hat V_{12}.
\end{equation}
The single-atom Hamiltonian $\hat{h}_k$ with respect to subsystem $k \in \{\mathrm{Ba}^+, \mathrm{Rb}\}$ is given by the sum of the kinetic energy operator ${\hat{T}_{\!k\!}}$ and the atomic binding potential ${\hat{V}_{\!k\!}}$ of the respective atom or cation.

In the cylindrical coordinates defined in Eq.~\eqref{eq:xBa}, 
the non-relativistic Jacobian-normalised kinetic energy operator for barium is expressed as
\begin{equation}
{\hat{T}_{\!\Ba\!}} := 
-\tfrac{\hbar^2}{2m} 
( {\partial_\zeta}^2 + {\partial_\rho}^2 )
-\tfrac{\hbar^2}{2m {{\rho}^{2}}} ({\partial_{\varphi_{\!\Ba}}}^{\!2} + \tfrac{1}{4})
\label{eq:TBa},
\end{equation}
where $\hbar$ is the reduced Planck constant, $m$ the electron rest mass, and $\partial_{x_j}$ is the shorthand for the partial derivative $\partial/(\partial x_j)$ with respect to coordinate $x_j \in \{\zeta, \rho, \varphi_{\!\Ba}\}$ such that e.~g.~$-i\hbar\partial_{\varphi_{\!\Ba}}$ corresponds to the canonical angular momentum operator along the cylindrical axis.
Due to the generalised structure of the program implementation, this kinetic energy operator is normalised with respect to the cylindrical Jacobian determinant as $\rho^{\frac{1}{2}}\,{\hat{T}'_{\!\Ba}}\,\rho^{-\frac12}$ where ${\hat{T}'_{\!\Ba}}$ represents the usual canonical kinetic energy operator. See the \autoref{a:jacobian} for further details, and \autoref{a:cylindrical} for the cylindrical case in particular. This method of simultaneous normalisation of wavefunction and kinetic operators in curvilinear coordinate systems is also called normalisation following the \textit{Dirac convention}.\autocite{nauts-mp1985}

The kinetic energy operator of the Rb atom in spherical polar coordinates is given by 
\begin{equation}
{\hat{T}_{\Rb} }=
-\tfrac{\hbar^2}{2 m} \,{\partial_r}^2 
+\tfrac{\hbar^2}{2 m{{r}^{2}}} \,{\hat{\ell}}^2_{({\theta},{\varphi_{\Rb}})},
\label{eq:TRb}
\end{equation}
where $\partial_r$ is the partial derivative with respect to the radial coordinate $r$ and $\hat{\ell}$ is the angular momentum operator.
The kinetic energy operator is normalised with respect to $r^2$, the Jacobian determinant as $r \,\hat T'_{\!\Rb} \,r^{-1}$ where $\hat T'_{\!\Rb}$ is the usual canonical kinetic energy operator. For further details refer to appendix \autoref{a:spherical}.
\paragraph{}
As Ba$^{+}$ and Rb are heavy elements with a single outermost valence electron,
we assume that we can describe them effectively hydrogen-like by a pseudopotential of frozen Ba$^{2+}$/ Rb$^+$ cores incorporating all tightly bound electrons on either subsystem $k$:\autocite{fuentealba-jphysb1982,fuentealba-jphysb1985}:  
\begin{equation}
\frac{4\pi\varepsilon_0}{e^2} \hat V_{k}(r_{k}) = - \frac{Z_{k}}{r_{k}} 
- \frac{\alpha_k}{2\, r_{k}^{~4}} \left( 1 -\exp[ -(\tfrac{r_{k}}{a_k})^2] \right)^2 
+ B_{k} \,\exp[- (\tfrac{r_{k}}{b_k})^2]
\label{eq:VBa}.
\end{equation}
Here, $\varepsilon_0$ is the electric vacuum permittivity, $e$ is the elementary charge, and $r_k$ is the distance between the centre of the nucleus $k$ and the respective electron. $Z_{\Ba} = 2$ and $Z_{\Rb}=1$ represent the effective charge numbers, $\alpha_{\Ba}$ and $\alpha_{\Rb}$ are the dipole polarisabilities of the closed-shell Ba$^{2+}$ and Rb$^{+}$, $a_k$ is the effective polarisation radius, and $B_k$ is a relativistic correction factor with associated radius $b_k$. In the respective literature the last summand of equation \eqref{eq:VBa} is a series expansion in the azimuthal quantum number $\ell$ of the electronic orbital. 
Since we are focusing in this work on exploring the feasibility of an atomistic quantum-dynamical model for \textsc{icec}, we are omitting the inclusion of projectors on the angular momentum. For now we treat the pseudopotential as $\ell$-independent and use the parameters for the spherical $s$ orbitals also for states of higher $\ell$ quantum numbers. In our exploratory simulation, electronic states with equal principal quantum number $n$ will therefore be degenerate.
The experimental energy levels of the bound $s$-states of Ba$^+$ and Rb \autocite{curry-jpcrd2004,sansonetti-jpcrd2006} are, however, reproduced very well in this approach. 
\txtrev{For the purpose of this study at the current state of knowledge
we wish to point out that we are conducting a simple proof of concept. While more accurate pseudopotentials exist, \autocite{marinescu-pra1994,aymar-rmp1996} 
 the presented computational method is compatible with more elaborate effective potentials that introduce additional parameters.
 Our simple potential is enough to show the aspects that need development. 
 The current shortcomings are discussed in the results discussion in terms of method and code development, which appear to be more significant at this point than the chosen potential itself.}

The incident electron forms an electric dipole with the Ba$^{2+}$ cation described by $e {\vec{x}_\Ba}$, while the bound rubidium valence electron forms an electric dipole $e {\vec{x}_\Rb}$ with the Rb$^{+}$ core. Therefore, the long-range interaction between those two electrons that are well separated on their respective atomic sites, may be expressed by an electric dipole-dipole potential
\begin{equation}
\frac{4\pi\varepsilon_0}{e^2} \, \hat V_{12}({\vec{x}_\Ba}, {\vec{x}_\Rb}, \vec{\distance}) = \frac{{\vec{x}_{\Ba}} \cdot {\vec{x}_{\Rb}}}{\distance^3} - \frac3{\distance^3} \left( {\vec{x}_{\Ba}} \cdot \tfrac{\vec{\distance}}{\distance}   \right) \!\left( \tfrac{\vec{\distance}}{\distance} \cdot {\vec{x}_{\Rb}}  \right)
.
\end{equation}
\txtrev{This description is in line with aspects arising from physical expectations toward a long-range interaction between the cation and atom:
	\begin{enumerate}
		\item 
		\textcolor{black}{
			In the well-separated case where $\distance \gg |\vec x_k|$, the dipole-dipole interaction will provide the first significant order of the multipole expansion of the Coulomb interaction. 
		}
As for the overall net-effect for atom-ion dimers both in their ground state $s$ orbital, however, the lower order permanent interaction terms vanish including the permanent dipole-dipole interaction term. \autocite{tomza-rmp2019,israelachvili-2011}
		The subsystems should be independent systems in simultaneous ground states; hence separable in wave function.
		We only expect an interaction during the timeframe of the electronic transition, \textit{i.e.} through the non-vanishing transition-dipole moments.
\item 
		The lowest order permanent interaction term in the multipole expansion for well-separated atom-ion dimers is then an attractive induced polarisation of order $\distance^{-4}$ arising from the net charge of the ion which is of higher order than a dipole-dipole interaction.
		Beyond the induced polarisation arising from the ionic net charge, the lowest non-vanishing interaction terms would arise from long-range atom-atom interaction in form of the van-der-Waals interaction of order $\distance^{-6}$.\autocite{tomza-rmp2019,stone-2013,israelachvili-2011,hapka-rsc2017} 
Any of those permanent contributions are of higher order and decay faster with distance. At this point of the development with respect to the question of non-local energy transfer, these higher-order permanent terms seem therefore of less significance under the assumption of a stabilized system due external trapping forces determining the interatomic distance.
	\end{enumerate}}
	Note that the condition $\distance \gg |\Vec{x}_{\Ba}|$ is not necessarily fulfilled rigorously at every moment in time for our simulation at this point of the model development. Apart from this interaction, the two atomic subsystems are thus completely independent in our description.
Within the chosen local subsystems (cf.~\autoref{f:coordinates}) of cylindrical geometry for barium~(II) and spherical polar symmetry for rubidium with both nuclei aligned along the $z$ axis, this interaction potential can be written explicitly as
\begin{multline}
	\frac{ 4\pi\varepsilon \distance^3 }{e^2} \, \hat V_{12}({\vec{x}_\Ba}, {\vec{x}_\Rb}, \vec{\distance}) =  
    \\ \left[ 
       \left(       \rho \sin\varphi_\Ba  \cos\varphi_\Rb
       +   {\zeta }  \sin\varphi_\Rb \right)\sin\vartheta
	   - 2 {\rho  } \cos\varphi_\Ba \cos\vartheta  \; \right] r \sin\alpha
     \\+ \left[
             \rho \cos(\varphi_\Ba -\varphi_\Rb) \sin\vartheta 
      - 2 {\zeta} \cos\vartheta 
    \right]  r\cos\alpha \;
  \label{eq:V12}.
\end{multline}

\subsubsection{The Wavefunction}\label{s:theory-wavefunction}
\paragraph{}
We use the Born-Oppenheimer approximation and treat the nuclei as effectively stationary and the electrons dynamically due to the large mass and therefore velocity difference between the two nuclei and the electrons.
The internuclear position vector $\vec \distance$ is treated as a fixed parameter and does not enter into our wave function.
We model the electronic wave function with the \textsc{mctdh} method\autocite{meyer-chemphyslett-1990,manthe-jchemphys-1992,meyer2009-1} in the form of
\begin{equation}
\Psi(t,{\vec{x}_{\Ba}},{\vec{x}_{\Rb}}) =\hskip-1.5ex\sum_{\mathrm{J}:=({j_\zeta, \dots}, {j_r, \dots})} \hskip-2.5ex A^J_{(t)}{\left(
	\mathcal{Z}^{j_\zeta}_{(t,\zeta)} \mathcal{Q}^{j_\rho}_{(t,\rho)} \mathfrak{f}^{j_\varphi}_{(t,\varphi_{\!\Ba})} 
\right)} {\left(
	\mathfrak{u}^{j_r}_{(t, r)} Y^{j_Y}_{(t, \theta,\varphi_{\Rb})} 
\right)}, \label{eq:mctdh}
\end{equation}
where $t$ is time and $J$ is a tuple of indices $({j_\zeta, \dots}, {j_r, \dots})$ associated with each (spatial) electronic degree of freedom. The time-dependent functions $\mathcal{Z}^{j_\zeta}_{(t,\zeta)}$, $\mathcal{Q}^{j_\rho}_{(t,\rho)}$ and $\mathfrak{f}^{j_\varphi}_{(t,\varphi_{\!\Ba})}$ encode the wavefunction components with respect to the degrees of freedom of the incoming electron on the barium~(II) cation, while $\mathfrak{u}^{j_r}_{(t, r)}$ and $Y^{j_Y}_{(t, \theta,\varphi_{\Rb})}$ correspond to time-dependent functions with respect to the radial and the angular degrees of freedom of the electron associated with rubidium. 
The time-dependent coefficients $A^J$ thus entangle the configurations by linear superposition. Since there is no direct exchange of electrons between the atomic sites at typical spatial separation for ultracold atoms in optical lattices expected, nor governed by the current description through distinct local coordinate subsystems, we do not impose any spin symmetry across the two atoms.

\paragraph{}
The initial wave function $\Psi_0$ is given by a combination of a free Gaussian probability distribution approaching the barium~(II) cation along the $\zeta$ direction and the ground state of the rubidium atom:
\begin{equation}
	\Psi_0 :=  
	{
		\mathcal{Z}_{0}(\zeta) \,\mathcal{Q}_{0}(\rho) \, \mathfrak{f}_{0}(\varphi_\Ba)
	} \;{
		\mathfrak{u}_{0}(r) \, Y_{0}(\vartheta, \varphi_\Rb)
	} \;.
\end{equation}
Longitudinally, the incoming electron is initially described by the Gaussian probability distribution
\begin{equation}
	\mathcal{Z}_{0}(\zeta) := {(2\pi \Delta_\zeta^2)}^{-\frac14} \exp\left[-\frac{(\zeta-\zeta_0)^2}{4 \Delta_\zeta^2} + \frac i\hbar p_0 (\zeta-\zeta_0)\right], \label{eq:Z(zeta)}
\end{equation}
with spatial standard deviation $\Delta_\zeta$ initially centred at $\zeta_0$ and moving with group velocity $p_0/m$. This distribution in longitudinal momentum can be interpreted according to the Copenhagen convention as a statistical uncertainty in position, momentum, and consequently kinetic energy of the electron source. We thus calculate the process with a range of incoming energies as expected from quantum-mechanical and technical uncertainties in an experimental setup.
In the transverse directions, the free electron has a radial Gaussian distribution
\begin{equation}
\mathcal{Q}_{0}(\rho) := {(2\pi \Delta_\rho^2)}^{-\frac14} \; \sqrt{\rho} \,\exp\left[
	-\frac{\rho^2}{4 \Delta_\rho^2} 
\right]\label{eq:Q(rho)}
\end{equation}
of characteristic spatial width $\Delta_\rho$ and vanishing group velocity, normalised with respect to the cylindrical Jacobian determinant by an additional factor of $\sqrt{\rho}$. The electron probability distribution is assumed independent of the polar degree of freedom $\varphi_\Ba$ with
\begin{equation}
	\mathfrak{f}_{0}(\varphi_\Ba) := (2\pi)^{-\frac12} \;.
\end{equation}

\paragraph{}
The rubidium atom is initially in its ground state ($5s$), which is the solution of the radial stationary Schr\"odinger equation
\begin{equation}
\left(-\frac{\hbar^2}{2 m}\frac{\partial^2}{\partial r^2} 
 +  V_{\!\Rb}(r)\right) {u}_{n}(r) = E_{n} ~u_{n}(r) \label{eq:u(r)}
\end{equation}
with lowest energy. Here, $n$ is the principal quantum number and $u_{n}(r)$ the Jacobian-normalised radial wave functions of the $s$ orbitals with energy $E_n$. The $s$ wave functions are radially symmetric such that the angular part is given by the normalisation constant
\begin{equation}
	Y_{0}(\vartheta, \varphi_\Rb) := (4\pi)^{-\frac12} \,.
\end{equation}

\section{Computational Details}\label{s:compdetails}
The electron dynamics was calculated in the effective two-electron treatment as described in \autoref{s:model} with the {\textsc{mctdh}} method.   We used the Heidelberg implementation of \textsc{mctdh} version 8, release 5, revision 8.\autocite{mctdh85,beck2000-1}
The physical parameters to set-up the individual potentials and the initial wave function, as well as the computational parameters to implement the underlying basis functions in their discrete variable representations ({\textsc{dvr}})\autocite{beck2000-1,light1992-185,light2000-263} are given in \autoref{t:parameters}. 
The continuum character across three spatial dimensions is resource-demanding to cover the free space and necessary phase space. The \textsc{MCTDH} algorithm allows to use a significant amount of primitive basis functions to form the configuration space in a discrete variable representation but remains numerically operational by reducing the stored coefficient vector to a limited number of dominant configurations selected and propagated on the full configuration space of single-particle functions (\textsc{spf}).\autocite{manthe-jchemphys-1992,meyer-chemphyslett-1990} Refer to Eq.~\eqref{eq:mctdh} for the wavefunction ansatz in particular, \autoref{s:theory-wavefunction} in general for the inital wavefunction definition and to \autoref{t:parameters} for the corresponding parameters.

\begin{table}[!h]\centering
	\caption[Computational Parameters for Atoms]{Collection of computational parameters used to model the scattering and capture of an incoming electron on Ba$^{2+}$ with a Rb atom at relative position $\Vec{\distance}$. The parameters are given in atomic units, where $\bohr$ represents the Bohr radius, $\hbar$ the reduced Planck constant, and $\hartree$ the Hartree energy.} \label{t:parameters}
	\renewcommand{\arraystretch}{1.1}\normalsize
	\renewcommand{\baselinestretch}{1.1}\normalsize
	\begin{tabular}{lrlr}
		\toprule
		\multicolumn{4}{l}{Frozen-core pseudopotential parameters,\autocite{fuentealba-jphysb1982,fuentealba-jphysb1985} see Eq.~\eqref{eq:VBa} }\\
		$Z_{\Ba}=2$ &\hspace*{-1em} $\alpha_{\Ba}=10.17~\bohr[3]~~\;$ & \multicolumn{2}{c}{$a_{\Ba}=~\,2.06~\bohr$\hspace*{1em}}  \\
		\multicolumn{2}{r}{$B_{\Ba}=16.71~\bohr[-1]~\,$}&\multicolumn{2}{c}{$b_{\Ba}=~\,1.2543~\bohr$}\\
$Z_{\Rb}~~=1$ &\hspace*{-1em} $\alpha_{\Rb}\;=~\,8.67~\bohr[3]~~~$ &\multicolumn{2}{c}{ $a_{\Rb}\,=~\,2.09~\bohr$\hspace*{1.1em}}  \\
		\multicolumn{2}{r}{$B_{\Rb}\;=45.272~\bohr[-1]$}& \multicolumn{2}{c}{$b_{\Rb}\,=~\,0.9941~\bohr$}\\
        \midrule 
        
		Interatomic parameters, see Eq.~\eqref{eq:V12} &{$\alpha = 90^{\circ}$\hspace*{2.5em}}&\multicolumn{2}{c}{$\distance~~\,=50.0~\bohr$\hspace*{1.25em}}\\
\midrule
		
		\multicolumn{4}{l}{Incident electron parameters, see Eqs.~\eqref{eq:Z(zeta)} and \eqref{eq:Q(rho)}}\\
		$p_0 = 0.467~{\hbar}/{\bohr}$ & $ \zeta_0= -85.0~\bohr$ & $\Delta_\zeta = 10.0\,\bohr$ & $\Delta_\rho = 10.0\,\bohr$\\ 
\midrule
		
		{\textsc{dvr} \hspace*{5.25em}type} & {grid points} & \multicolumn{2}{c}{ range }   \\
$\zeta$    \hspace*{6.5em} \textsc{fft}  &  243 & $-156.50~\bohr$&$+156.50~\bohr$ \\
		$\rho$    \hspace*{2.25em}Generalised Laguerre $L^{(1)}_n$\hspace*{-5em}  &  125   &  \hskip1.7em$0.01~\bohr$&$159.77~\bohr$ \\
		$\varphi_{\!\Ba}$     \hspace*{1.5em}Periodic Exponential\hspace*{-5em}    & 15  & \hspace*{1.7em}$0$&$2\pi$\hspace*{1.2em} \\
		$r$   \hspace*{2.25em}Generalised Laguerre $L^{(2)}_n$   \hspace*{-5em}& 125 &  \hskip1.7em$0.02~\bohr$&$~160.43~\bohr$  \\
		$(\vartheta,\varphi_{\!\Rb})$  \hspace*{0.5em}Extended Legendre\hspace*{-5em}   &   7  & \multicolumn{2}{c}{ ~~\,$\ell_\Rb \in \{0, ~\ldots, ~~~\,6\}$ } \\
		&   7 &  \multicolumn{2}{c}{ \,$m_{\ell_\Rb} \in \{0, ~\ldots, ~\pm3\}$ } \\

		\midrule
		\multicolumn{1}{l}{\textsc{spf} configuration space $(\dim A^J_{(t)})$, see Eq.~\eqref{eq:mctdh}\hspace*{-4em}}&\multicolumn{3}{c}{$(12 {\times} 12 {\times}  7) ~\times~ (12 {\times} 12)$}\\

		\midrule
		Complex absorbing potential, see Eq.~\eqref{eq:cap} &$n$\hspace*{1.5em}  $\kappa$ &\multicolumn{1}{c}{$z_{cap}$} & \multicolumn{1}{c}{$\eta$}  \\ 
		&$2$ ~\, $\pm1$ &$\pm100.0~\bohr$& \hspace*{-1em}$52.0\times10^{-6}~\hartree$   \\

		\bottomrule
	\end{tabular}
\end{table}

\paragraph{}
The parameters for the atomic pseudo\-potentials have been taken from the available lit\-era\-ture.\autocite{fuentealba-jphysb1982,fuentealba-jphysb1985} 
The interatomic distance $\distance$ between the centres of the two species is set to $50.0~\bohr$. This choice is a compromise at this stage of the model development between usually larger interatomic distances in comparable experiments for optically trapped atoms\autocite{bahrami-physstatsol2019,kruekow-physrevA2016} and the fast decline of the dipole-dipole interaction with $\distance^{-3}$. The angle between the axis $\vec \distance$ connecting the two atoms and the electron beam axis $\zeta$ is taken to be $\alpha = 90^{\circ}$. The impact of changing this parameter should be investigated in the future if the model appears viable at this stage as will be indicated in this work.

\paragraph{}
The initial group momentum $p_0$ of the incoming electron is set to $0.467~\hbar/\bohr\ $ (2.97 eV). Since this energy lays well under the ionisation potential of Rb, the capture in an excited state of Ba$^+$ with subsequent excitation of the Rb atom is energetically allowed. 
\txtrev{The individual subsystems themselves, and therefore atomic trap energies operate at a few microkelvin only (up to 1 mK).
However, the overall atom-ion dimer is in an excited state of about 5 eV with respect to the overall isoelectronic ground state,\autocite{NIST:ASD} which would be Ba$^+$--$Rb^+$.
}
The electron's initial longitudinal position is $\zeta_0 =-85.0~\bohr$; the longitudinal and lateral initial widths are $\Delta_{\zeta/\rho}= 10~\bohr$. This corresponds to a width of $\Delta_{\displaystyle\epsilon} = 0.64~\eV$ in the energy distribution. These parameters for the initial wavefunction were chosen to mediate a few numerically limiting aspects. First, the wavepacket needs to have a reasonable spread in position space such that it is bigger than the size of a low-energy bound atomic orbital, while its associated normal distribution in momentum space should be reasonably broad yet sufficiently contained. This is required for the incoming electron to have low enough energy to avoid ionising diffraction and so that its wavelength is not of the same order of magnitude as the atom size. The slower fractions in the momentum distribution should still be travelling towards the barium~(II) cation and should be fast enough for the computational simulations to be completed within a reasonable amount of time steps. 
Lastly, the parameters are meant to be reasonable for cold atom experiments. However, we note that these numbers are not fully optimised for reaction probabilities.

\paragraph{}
Different types of discrete variable representations (\textsc{dvr}) were chosen for the representation of the wavefunction with respect to the individual electron coordinates in accordance with their geometric properties and computational performance, cf.~\autoref{s:theory-wavefunction}. The ranges in real space and the amount of grid points have been balanced between a necessary coverage of physical space, a grid-point density to cover necessary position- and momentum-space, and a limit on total grid size that is still computable. The longitudinal $\zeta$ direction is modelled by a Fast-Fourier-Transform \textsc{dvr} of 243 grid points, the transverse counterpart $\rho$ by 125 grid points on a generalised Laguerre $L_n^{(1)}$ \textsc{dvr} which matches the circular boundary conditions with 15 grid points on the periodic exponential \textsc{dvr} in the angular $\varphi_{\!\Ba}$ direction. As the incident electron has a cylindrical geometry while the atomic pseudopotential has spherical symmetry, the \texttt{potfit} algorithm of \textsc{mctdh} needed to be employed to transform the binding potential of Eq.~\eqref{eq:VBa} into an expansion in products of $\zeta$ and $\rho$.
On the spherical subsystem of rubidium, the radial degree of freedom $r$ is handled by 125 grid points on a generalised Laguerre $L_n^{(2)}$ \textsc{dvr} and the angular components on an extended Legendre \textsc{dvr} known as \texttt{KLeg} by 7 angular momentum quantum numbers $\ell_\Rb$ and 7 magnetic quantum numbers $m_{\ell_\Rb}$ between -3 and +3. The chosen generalised Laguerre \textsc{dvr} $L_n^{(1)}$ match the Jacobian determinant factor of $\rho$ for the cylindrical radial coordinates, while $L_n^{(2)}$ match the Jacobian determinant factor of $r^2$ for the spherical radial coordinate. 
Twelve single-particle functions (\textsc{spf}) per \textsc{dvr} are combined, except for seven in $\varphi_{\!\Ba}$ to form the dimension of the time-dependent coefficient tensor $(A^J_{(t)})_J$ to the wavefunction according to Eq.~\eqref{eq:mctdh}. With those parameters, the simulation accounts for 225 bound states in Ba$^+$ and 91 in Rb. For a detailed list of parameters, see \autoref{t:parameters}, and for an appropriate description of computational aspects in more depth, see the \textsc{mctdh} references.\autocite{mctdh85, meyer2009-1,  beck2000-1, manthe-jchemphys-1992, meyer-chemphyslett-1990} 

\paragraph{}
To prevent non-physical reflections of electron probability density at the edges of the simulated volumes we add a complex absorbing potential (\textsc{cap}) at the boundaries marked schematically by $\vol_{Ba}$ and $\vol_{Rb}$ in \autoref{f:coordinates}. 
We thus treat electrons that are too far away from the Ba$^{2+}$ and Rb$^+$ cores as scattered and ionised, respectively. 
The \textsc{cap}s are given by:
\begin{equation}
	-i \eta \, [ \kappa z - \kappa z_{cap} ]^n ~\Theta_{ (\kappa z -\kappa z_{cap}) }\,,
 \label{eq:cap}
\end{equation}
where $\eta$ is the strength of the complex absorbing potential in units of energy, $\kappa$ is typically $\pm1$, $z$ stands for the respective continuum direction $z\in \{\zeta ,  \rho , r\}$ here, and $\Theta_{(z)}$ represents the Heaviside function as function of $z$. 
A \textsc{cap} onset position of $z_{cap}$ at $\pm100~\bohr$ was chosen to allow extended bound-state wave functions. An extended stationary wavefunction is expected for highly excited bound states like Rydberg states. The potential strength $\eta$ was carefully adjusted to the range of energies involved in the computation and the effective length available to the complex absorbing potential, see \autoref{t:parameters} for a summary of these details. For this system where each direction was covered to around $160~\bohr$, the \textsc{cap} strength $\eta$ of $52 \times 10^{-6}~\hartree$  was found appropriate for the present range of continuum energies.

The introduction of the \textsc{cap} into the simulation adds a non-Hermitian part to the Hamiltonian. This non-Hermitian constituent instigates a decay-like behaviour for those fractions of the time-dependent wavefunction that are reaching beyond the respective $z_{cap}$.\autocite{beck2000-1,moiseyev-physrep1998} This decay within the wavefunction describing particles simulated within a numerical box, indicates the physical equivalent of particles leaving the simulated multi-dimensional volume. The two simulated electrons in our model are entangled in the wavefunction ansatz per configuration denoted by composite index $J$ and associated with the time-dependent coefficient $A^J_{(t)}$, cf. Eq.~\eqref{eq:mctdh}. Numerically, this implies once one of the two electrons passes into a \textsc{cap} in any coordinate, the respective configuration is annihilated from the total wave function. Therefore we also numerically lose the correlated wavefunction contribution from the other potentially still bound electron. This is a numerical feature that we expect from throughout previous quantum-dynamically investigated systems.\autocite{pont2019-prepare,molle-jcp-2019,Haller2019,pont-jphys-2016,bande-epjconf-2015,pont-physrev-2013,bande-jchemphys-2011} We will account for this numerical phenomenon in our interpretation of the simulation results presented in the following section.

\section{Results and Discussion}\label{s:results}

In order to observe the environment-assisted electron capture, we analyse the electron probability density in \autoref{s:results-DensEvol} as function of time $t$ and projected onto one spatial degree of freedom at a time by integrating over all other coordinates. 
We will focus on the longitudinal coordinate $\zeta$ and the transverse coordinate $\rho$ to trace the impinging electron and potentially its binding to, or its scattering from Ba$^{2+}$. Similarly, to be able to observe the emission or excitation of an electron from Rb we will investigate the projected electron probability density as a function of radial coordinate $r$.  
In \autoref{s:results-flux} complementary information is then gained by analysing the flux of electron probability densities through the respective \textsc{cap} surface as functions of time $t$. By analysing the data we can calculate reaction probabilities for the environment-assisted capture. We can distinguish between emission and excitation of the rubidium electron due to assisted electron capture at the barium~(II) site. This is presented in \autoref{s:reaction-prob}. 

\subsection{Probability Density Evolution}\label{s:results-DensEvol}

\begin{figure}[!h]\centering
\hspace*{-.1\textwidth}\tikz\node (int)
    {\includegraphics[width=0.34\textwidth]{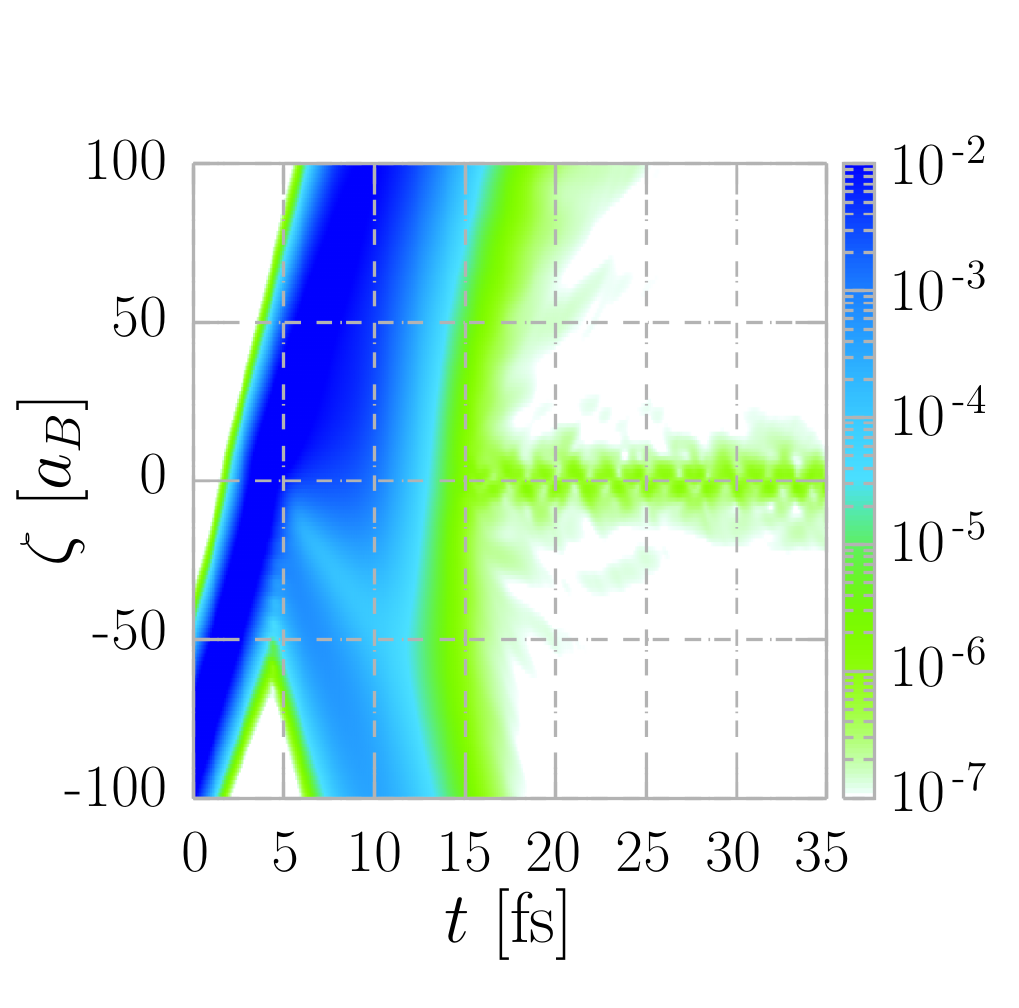}} 
    node[anchor=north west] at (int.north west) {a)}
    node[anchor=north] at (int.north) {Interaction}
    node[anchor=west, xshift=-.75em] (ref) at (int.east) 
    {\includegraphics[width=0.34\textwidth]{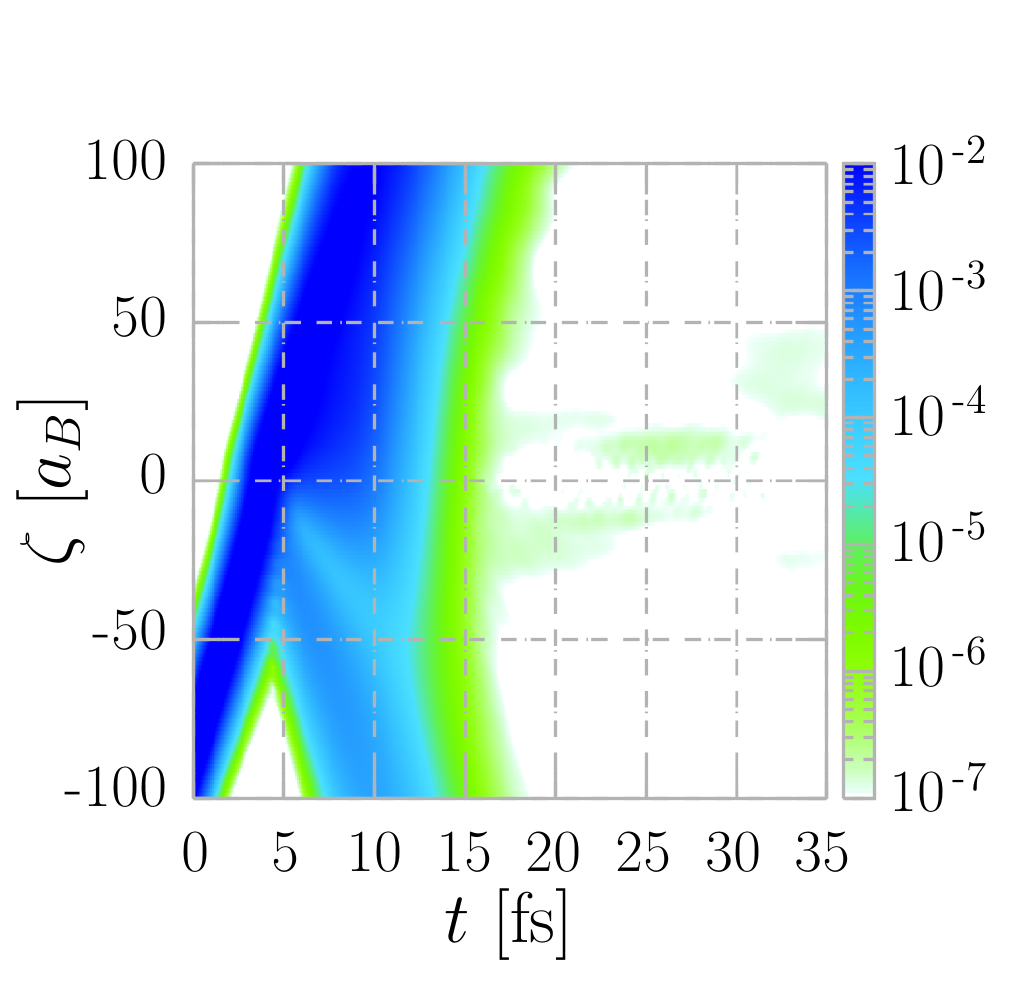}} 
    node[anchor=north west] at (ref.north west) {b)}
    node[anchor=north] at (ref.north) {No Interaction}
    node[anchor=west, xshift=-.75em] (dif) at (ref.east)
    {\includegraphics[width=0.34\textwidth]{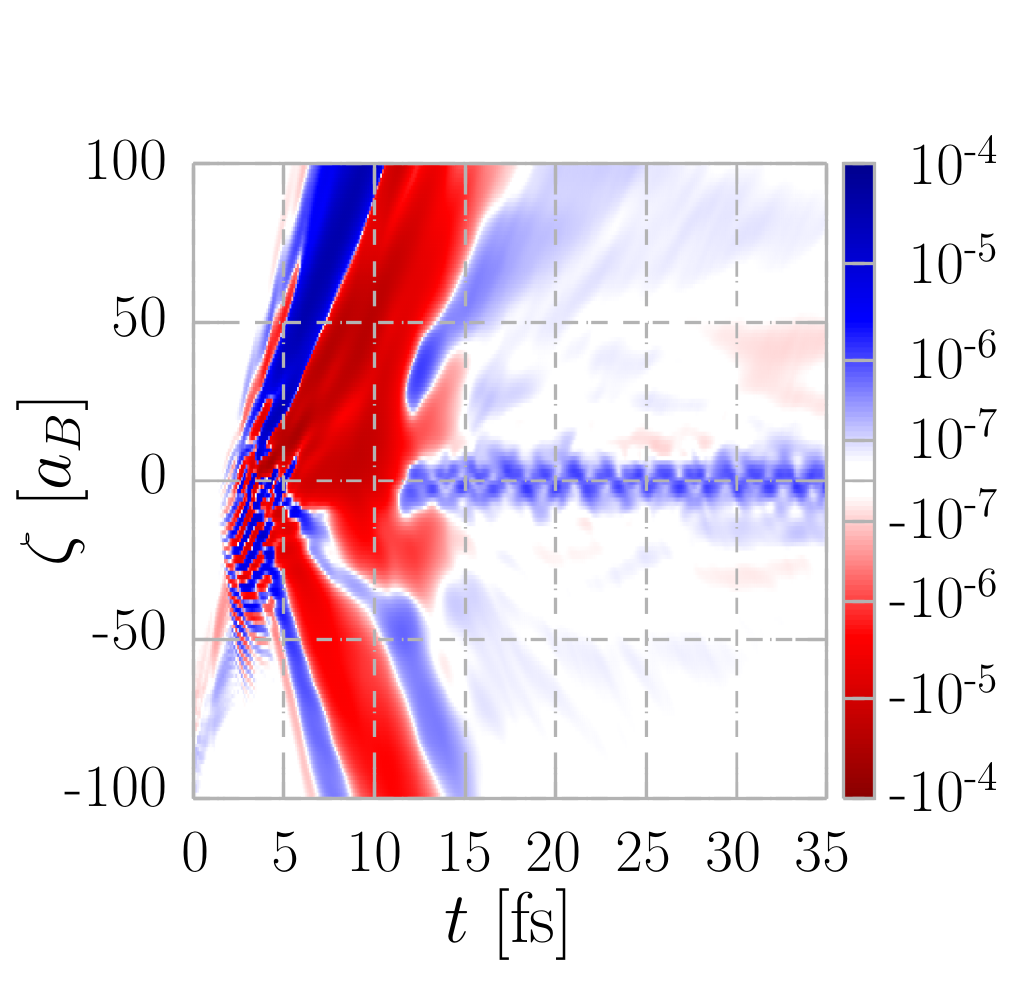}} 
    node[anchor=north west] at (dif.north west) {c)}
	node[anchor=north] at (dif.north) {Difference};
	\hspace*{-.1\textwidth}
\caption{Evolution of the electron probability density as a function of the longitudinal electron-beam coordinate $\zeta$ and time $t$. The barium~(II) core is situated at the origin, $\zeta=0$. The probability density is normalised to unity at $t = 0~\fs$ and the logarithmic colour axis reflects its absolute value. Plot a) displays the dynamics with the full Hamiltonian, plot b) shows a simulation with the dipole-dipole interaction turned off, representing the scattering at an isolated barium~(II) cation. Plot c) indicates the difference between the two simulations in symmetrised logarithmic (symlog) colour. 
	}\label{f:f1}
\end{figure}

In \autoref{f:f1} the electron probability density $|\Psi(t,\zeta)|^2$ is shown as a function of time $t$ and the longitudinal coordinate $\zeta$ to trace the impinging electron and potentially its binding to Ba$^{2+}$ or its scattering. The density is normalised as $\int |\Psi(t=0,\zeta)|^2 d\zeta =1$. In plot a) the two subsystems are able to exchange energy through dipole-dipole interaction, in contrast to plot b) where energy exchange across the systems is disabled for the simulation.
Within about the first $4~\fs$, only the motion of the initially unbound electron towards the Ba$^{2+}$ ion at position $\zeta=0$ is visible. During that time, the Gaussian-shaped wave packet slightly widens while its intensity gradually sinks, which is the diffluence of density common to all freely propagating wave packets. Once the electron reaches Ba$^{2+}$, scattering occurs. It induces a quick dispersion of the wave packet over the entire available longitudinal domain, where a larger part of the wave packet surpasses the ion while the rest is scattered back. When the front fractions of the wave function are backscattered, they encounter the tail fractions still moving toward the cation which leads to self-interference and is depicted as a brighter region in the probability density propagating from $(t,\zeta)=(5~\fs,-15~\bohr)$ to $(t,\zeta)=(10~\fs,-45~\bohr)$. 

\begin{figure}[!h]\centering
\hspace*{-.1\textwidth}\tikz\node (int)
    {\includegraphics[width=0.34\textwidth]{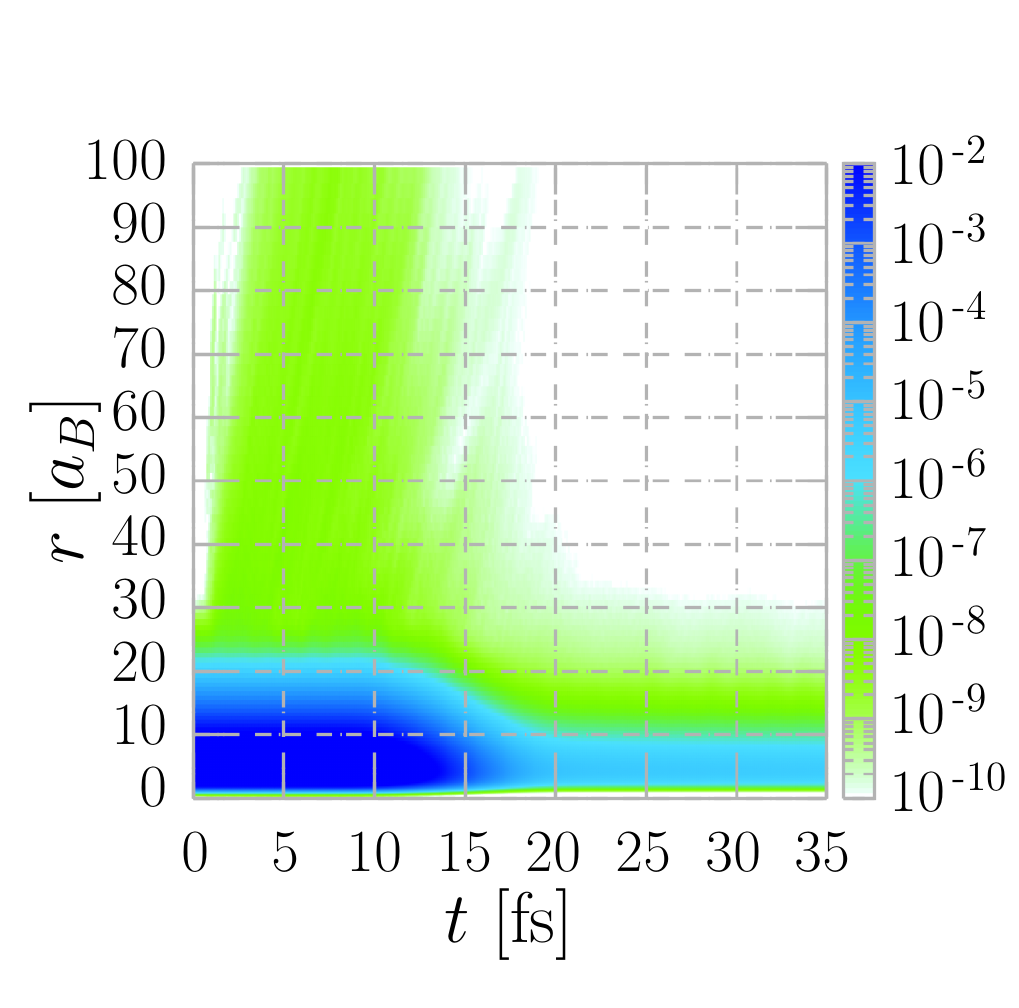}} 
    node[anchor=north west] at (int.north west) {a)}
    node[anchor=north] at (int.north) {Interaction}
    node[anchor=west, xshift=-.75em] (ref) at (int.east) 
    {\includegraphics[width=0.34\textwidth]{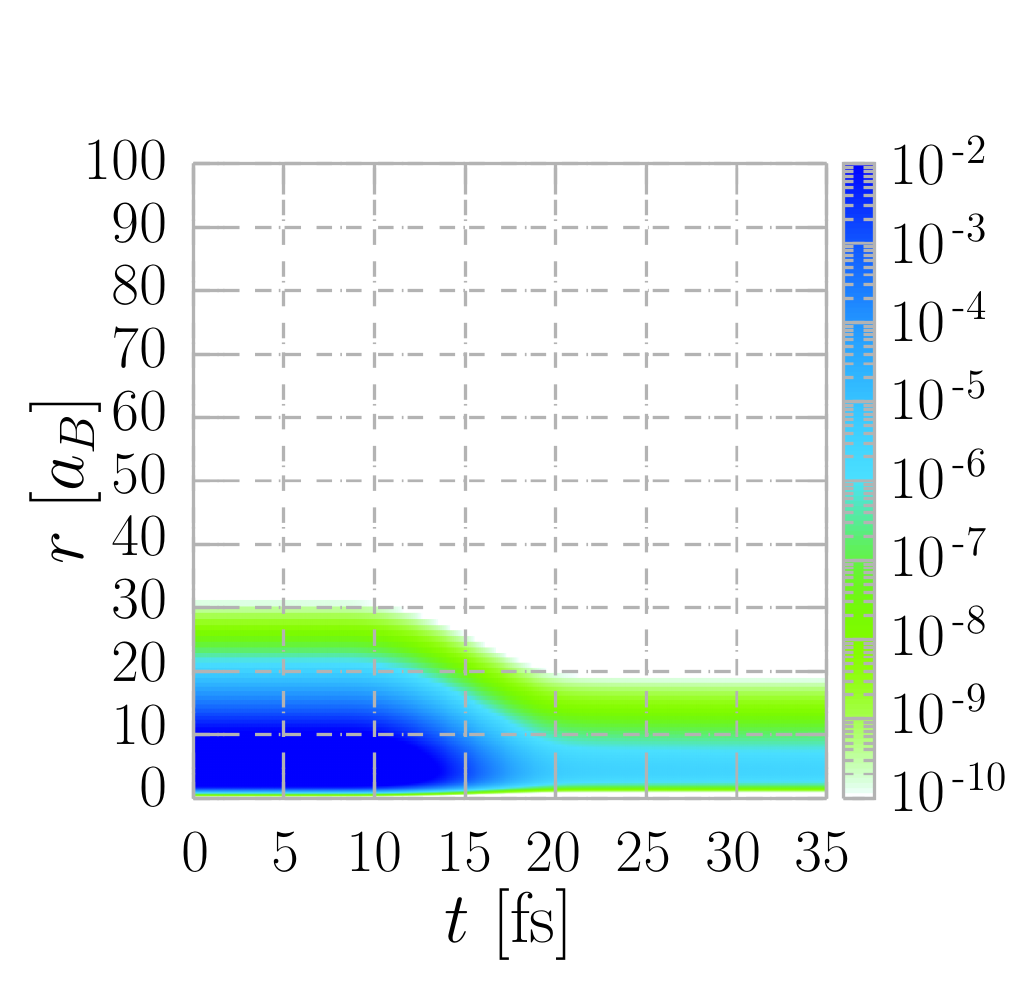}} 
    node[anchor=north west] at (ref.north west) {b)}
    node[anchor=north] at (ref.north) {No Interaction}
    node[anchor=west, xshift=-.75em] (dif) at (ref.east)
    {\includegraphics[width=0.34\textwidth]{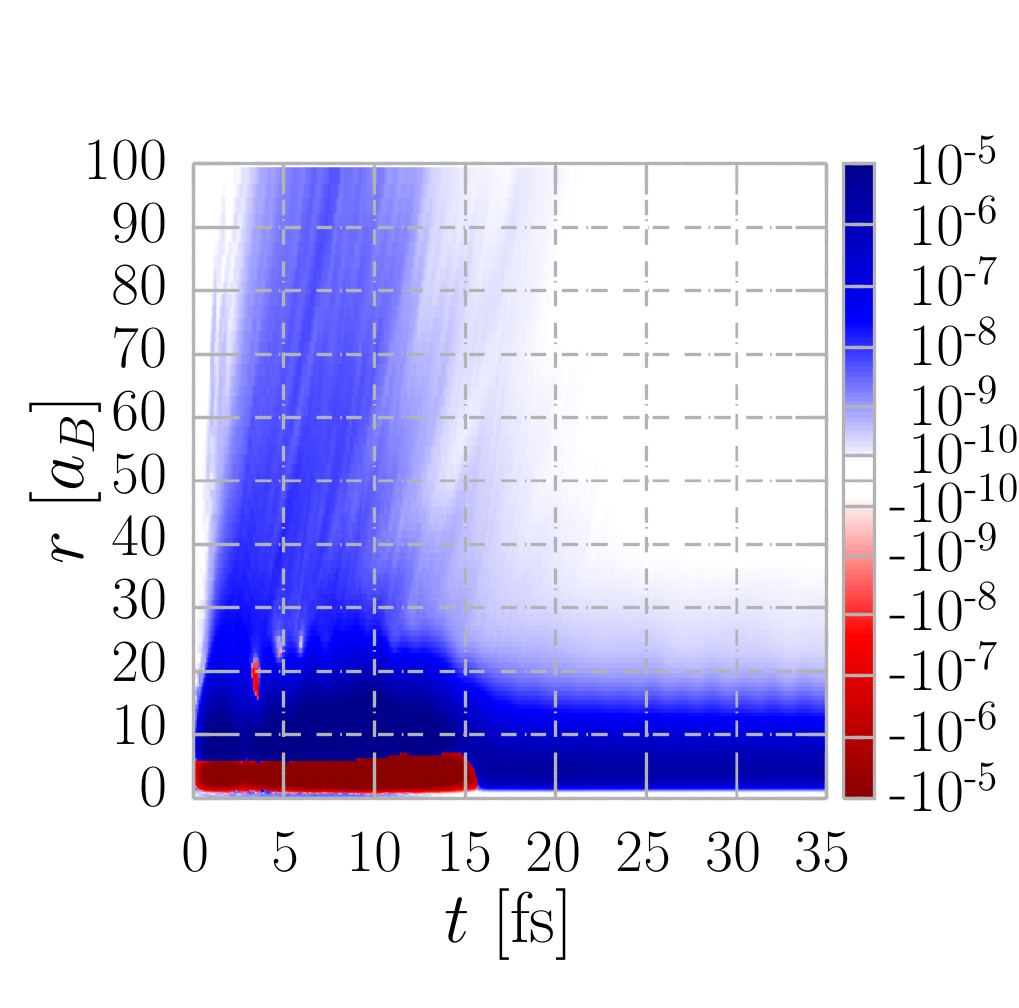}} 
    node[anchor=north west] at (dif.north west) {c)}
	node[anchor=north] at (dif.north) {Difference};
	\hspace*{-.1\textwidth}
\caption{Evolution of the absolute electron probability density as function of the radial position $r$ in the rubidium subsystem and of time $t$. Plot a) shows additional effects on the outer valence electron of rubidium due to interatomic energy transfer, plot b) the simulation for the entangled wavefunction without interatomic interaction, and c) the difference between these two.
    The probability density represented as time-dependent function of the rubidium subsystem declines considerably between $10$ and $20~\fs$ for both a) and b), due the electron diffraction at barium and the entangled wavefunction across the subsystems according to Eq.~\eqref{eq:mctdh}. The surplus of localised probability density in plot c) after $16~\fs$ indicates therefore a bound electron at Rb correlating with a captured electron at Ba.
	}\label{f:f4}
\end{figure}

After about $15~\fs$ almost all incident probability density has scattered at barium and reached the respective \textsc{cap} where it is absorbed. This becomes apparent from the strong overall reduction of the norm. Electron probability density is only visibly retained in the region $\zeta \in [-10, 10]~\bohr$. Since this bit of electron probability density of the order of $10^{-6}~\bohr[-1]$ only remains near the scattering center if the dipole-dipole interaction is turned on such that the release of excess energy is possible (compare to plot b)), this indicates successful environment-assisted electron capture. This is most apparent in plot c), which shows the difference between the cases when the dipole-dipole interaction is on or off. When the interaction is turned off, or physically Rb is at infinite distance to Ba$^{2+}$, the electron cannot be captured as there is no way to transfer energy out of the barium subsystem. Note however that we only see the captured electron density at barium if the rubidium electron does not reach its respective \textsc{cap} (cf.~discussion in context of \autoref{f:f4}, and an overview of all time-dependent results together in \autoref*{a:summary},\autoref{f:summary}). This is due to the entangled nature of the two electrons in the wave function which leads to that fraction of the full two-electron wave function getting annihilated when one of the electrons reaches its respective \textsc{cap}. This means that the remaining probability density we observe at barium corresponds to capture with excitation or low energy emission of the rubidium electron. Although not perfectly visible in the graphical representation, the captured density slowly reduces over time indicating that the electron probability density is gradually emitted from rubidium towards its \textsc{cap}, or that the temporarily captured probability density at barium gets re-emitted gradually even for $t > 20~\fs$. Furthermore, we see an oscillation in the captured electron density due to a superposition of the bound states.  

\begin{figure}[!h]\centering
\hspace*{-.1\textwidth}\tikz\node (int)
    {\includegraphics[width=0.34\textwidth]{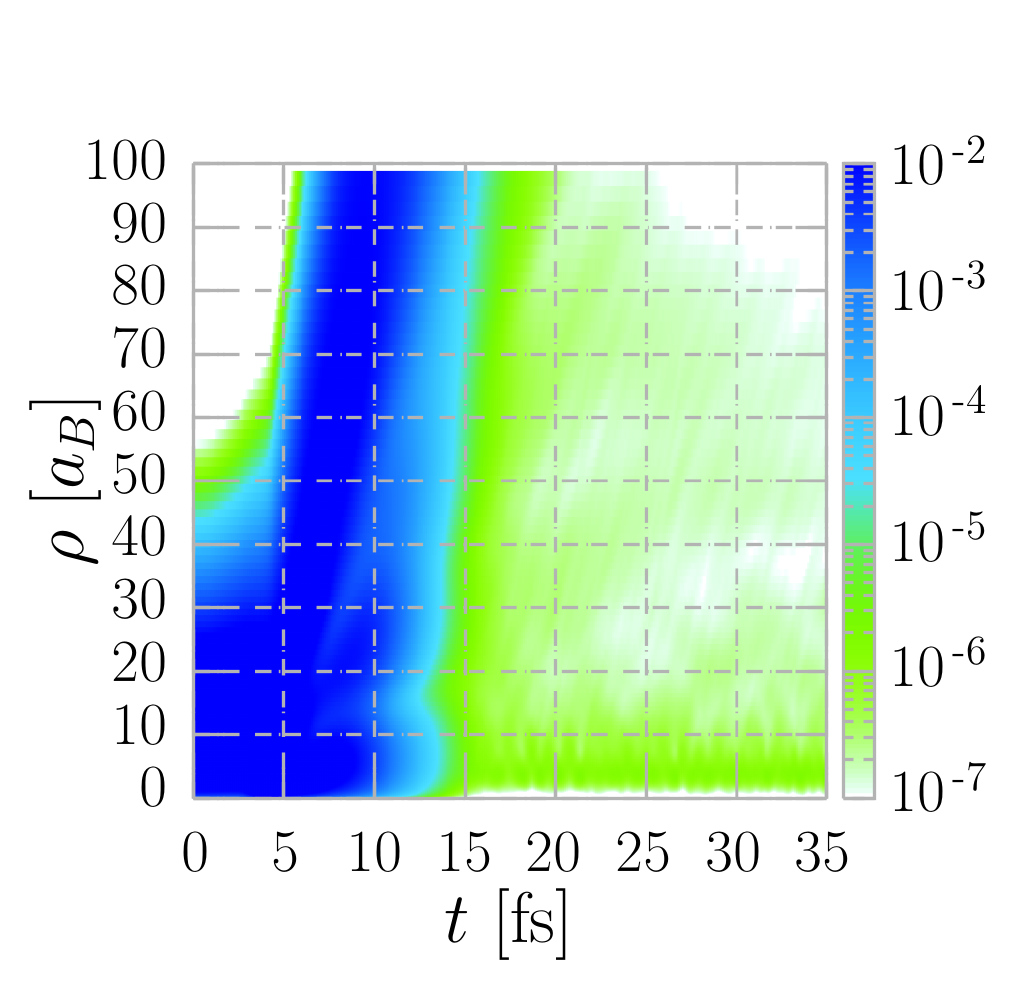}} 
    node[anchor=north west] at (int.north west) {a)}
    node[anchor=north] at (int.north) {Interaction}
    node[anchor=west, xshift=-.75em] (ref) at (int.east) 
    {\includegraphics[width=0.34\textwidth]{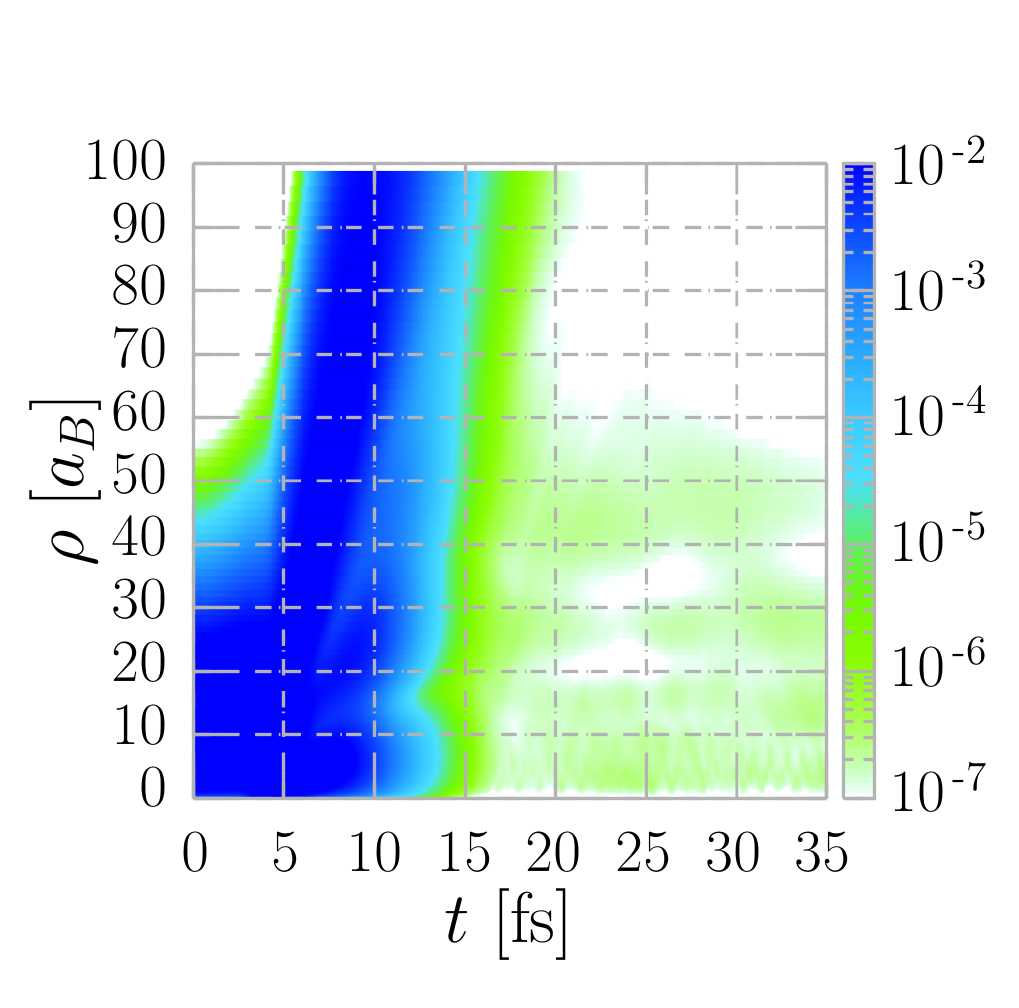}} 
    node[anchor=north west] at (ref.north west) {b)}
    node[anchor=north] at (ref.north) {No Interaction}
    node[anchor=west, xshift=-.75em] (dif) at (ref.east)
    {\includegraphics[width=0.34\textwidth]{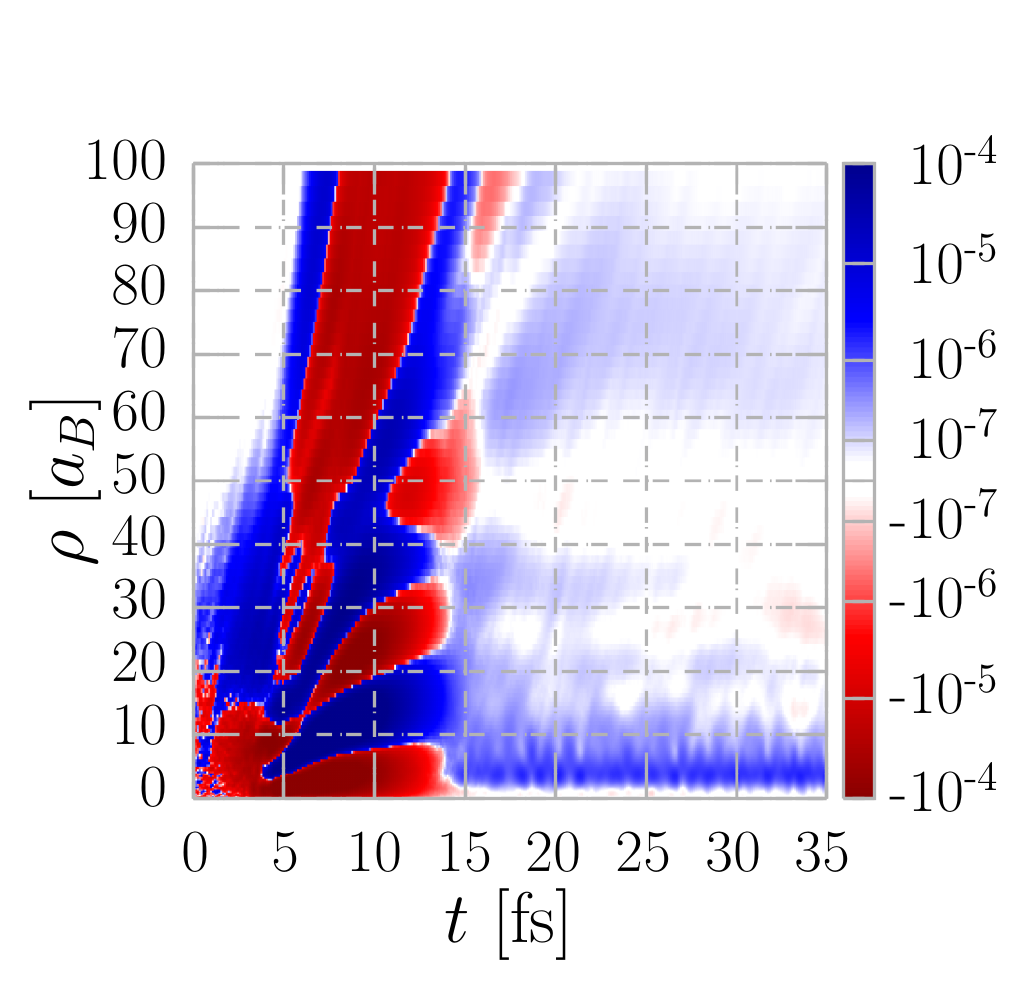}} 
    node[anchor=north west] at (dif.north west) {c)}
	node[anchor=north] at (dif.north) {Difference};
	\hspace*{-.1\textwidth}
\caption{Evolution of the electron probability density as a function of the transverse coordinate $\rho$ and time $t$. The probability is normalised to unity at $t = 0~\fs$ and the logarithmic colour axis reflects the absolute value of the probability density at a point in space. Plot a) presents the dynamics with the full Hamiltonian, plot b) represents a simulation without interatomic interaction, representing the scattering at an isolated barium~(II) cation. Plot c) shows the difference between the two simulations in symmetrised logarithmic (symlog) colour.
}\label{f:f2}
\end{figure}

\autoref{f:f2} a) shows the time evolution of the electron probability density after integrating over all coordinates but $\rho$. \txtrev{The coordinate $\rho$ represents the impact parameter in the traditional Rutherford scattering case}. 
It is normalised as $\int |\Psi(t=0,\rho)|^2 d\rho =1$. We also provide a collective graphical overview for the comparison of all time-dependent results in \autoref*{a:summary},\autoref{f:summary}.
The incoming wave packet initially has a maximum at $\rho=10~\bohr$ which gets focused while approaching the positively charged Ba$^{2+}$ core. At the same time the distribution becomes more dispersed and tails more extensively towards larger $\rho$ prior to the scattering due to natural broadening.

After the scattering event the probability density is dispersed quickly until at $t=15~\fs$ the majority of the wave packet reaches the \textsc{cap} where it is absorbed leading to a strong decrease of overall density. As for the longitudinal coordinate, also along the transverse coordinate, a small remainder of probability density at the order of $10^{-6}~\bohr[-1]$ establishes below $10~\bohr$ from the cylindrical axis. Again, this successful capture only appears if mediated by a dipole-dipole interaction, as seen in comparison to plot b) where the interaction disabled and in plot c) of \autoref{f:f2}, which shows the difference between these two simulations. 

It is worth noting that in the long-time regime ($t > 15~\fs$), bursts of constant momentum from the barium axis outward enter in the probability density. This can be best seen in the difference in plot c) at $t=20~\fs$ and at around $\rho=30~\bohr$ or $\rho=70~\bohr$. This continuous density emission could be a trace of a temporary capture in a resonance state before the electron is re-emitted after reabsorbing the excess energy from rubidium.

Lastly, the evolution of the electron initially bound to rubidium should be investigated, as environment-assisted electron capture processes involve the emission or the excitation of this electron.
\autoref{f:f4} shows the probability density at time $t$ after integrating over all coordinates but the radial position $r$ of the electron with respect to the Rb nucleus. It is normalised as $\int |\Psi(t=0,r)|^2 dr =1$.

The initial radial probability density has a maximum at $r = 4.4~\bohr$ and reaches up to about $r = 30~\bohr$. In the fully interacting case, almost immediately after the start of the simulation, electron probability density begins to spike out from the rubidium core. This shows the emission of electrons due to the transfer of energy from the barium to the rubidium subsystem. Similarly to the small amount of captured probability density observed in the discussion of the barium subsystem (cf.~\autoref{f:f1} and \autoref{f:f2}), also only a small amount of density gets emitted from rubidium. 

The very early emission of electron probability density after about $1~\fs$ can be explained as follows. 
The experimental ionisation threshold of Ba$^+$ is $10.00~\eV$,\autocite{NIST:ASD,karlsson-physscr1999} and that of Rb is $4.18~\eV$.\autocite{NIST:ASD,johansson-arkfys1961,lorenzen-physscr1983} This implies that direct \textsc{icec} into the ground state \txtrev{releases excess energy}. It will emit the rubidium electron with $5.82~\eV$ higher kinetic energy than the incident electron on barium.
The incoming electronic wave packet contains a range of energies coming into the system which is centred at about $3~\eV$, which is less than the ionisation potential of Rb. A near-threshold ionisation of rubidium by energy transfer would thereby capture the incident electron at group velocity already $1.18~\eV$ below the ionisation threshold of barium~(II). This represents the experimental potential energy of the $10p$ orbital.\autocite{NIST:ASD, sansonetti-jpcrd2006} 
A capture into an excited state at barium thereby suffices to transfer enough energy to the rubidium subsystem to ionise it. Moreover, the kinetic energy of 3~eV alone is enough to raise the rubidium electron from the $5s$ orbital to the $6p$ orbital.\autocite{NIST:ASD, sansonetti-jpcrd2006} The barium electron is hereby decelerated close to $0~\eV$ kinetic energy. 
The next higher excited rubidium orbital, $5d$, consumes already enough transferred energy to attach the incoming electron to the energetically high and spatially extended $21s$ orbital of Ba$^{+}$.\autocite{NIST:ASD, sansonetti-jpcrd2006,curry-jpcrd2004} With the continuous distribution of incident energy, the higher-lying Rydberg states of Ba$^{+}$ are also already available as capturing states. As these weakly bound states of Ba$^+$ extend far from the atomic core on the microscopic scale, the incoming electron can be captured into them very early. That early capture would be hidden in the overall electron probability distribution in the barium coordinates, as shown in \autoref{f:f1} and \autoref{f:f2}, because it is dominated by the elastic (Rutherford) scattering of the electron from the barium core. Furthermore, \autoref{f:f1} cannot distinguish between free electrons and electrons captured in a Rydberg state. We can thus only see the beginning of the capture process translating into excitation and ionisation of the rubidium electron probability density shown in \autoref{f:f4}.

The majority of electron probability density with respect to the rubidium coordinate $r$ stays unaltered until at about $t= 15~\fs$, when there is a considerable decrease in the absolute electron probability density for both cases of simulation with interacting and isolated subsystems (compare plots a) and b) of \autoref{f:f4}). This is numerically due to the entanglement of the wavefunction across the subsystems which numerically removes the contributions that correspond to electron probability density in the barium subsystem that reach the respective \textsc{cap} after diffraction. This does not have any physical implications however and only indicates the portions of wavefunction that contain a free uncaptured barium electron. A renormalisation of the wavefunction would be possible to correct for this numerical feature. We have refrained from doing so because the electron diffraction is by orders of magnitude dominant over the interatomic processes we are interested in. 
At $15~\fs$, there is only 0.65\% of the initial two-electron wavefunction left simultaneously in the simulated subsystem contained in volumes $\vol_\Ba$ and $\vol_\Rb$.
At $18~\fs$, there are only 0.01\% left that have not been consumed as free electron portions by the \textsc{cap}.
A renormalisation compensating visually for those diffracted portions would therefore strongly overcompensate the representation and mask the interesting portions of the wavefunction in \autoref{f:f4}.
In the difference between the absolute probability density for the interactive and isolated case, we can see that the emission and excitation of the rubidium electron is due to the interatomic interaction (cf.~plot c) of \autoref{f:f4}). This gives us the last piece of evidence for successful environment-assisted electron capture.
Furthermore, we observe that the ionised fractions of probability density leave the rubidium core with different velocities, indicated by variations in slope on the $r$-$t$ plot. This is due to the continuous energy distribution of our incident wave packet on barium whose assisted capture into the whole band of discrete atomic orbitals results in a distribution of outgoing velocities for the ionising rubidium electron.

\txtrev{Typical atomic trap frequencies are $\sim$ 10~kHz, translating to
timescales of about 100~$\mu$s.
Considering the reaction timescale of $10~\fs$ on which \textsc{icec} happens, 
the system is nearly static during the reaction.
This justifies our model in which nuclei are static for the time of the simulation and during the reaction itself.
An experimental observation may therefore be possible within the stable timeframe of the system.}

\subsection{Flux Density}\label{s:results-flux}

\begin{figure}[!h]\centering
\hspace*{-.1\textwidth}\tikz\node (int)
    {\includegraphics[width=0.33\textwidth]{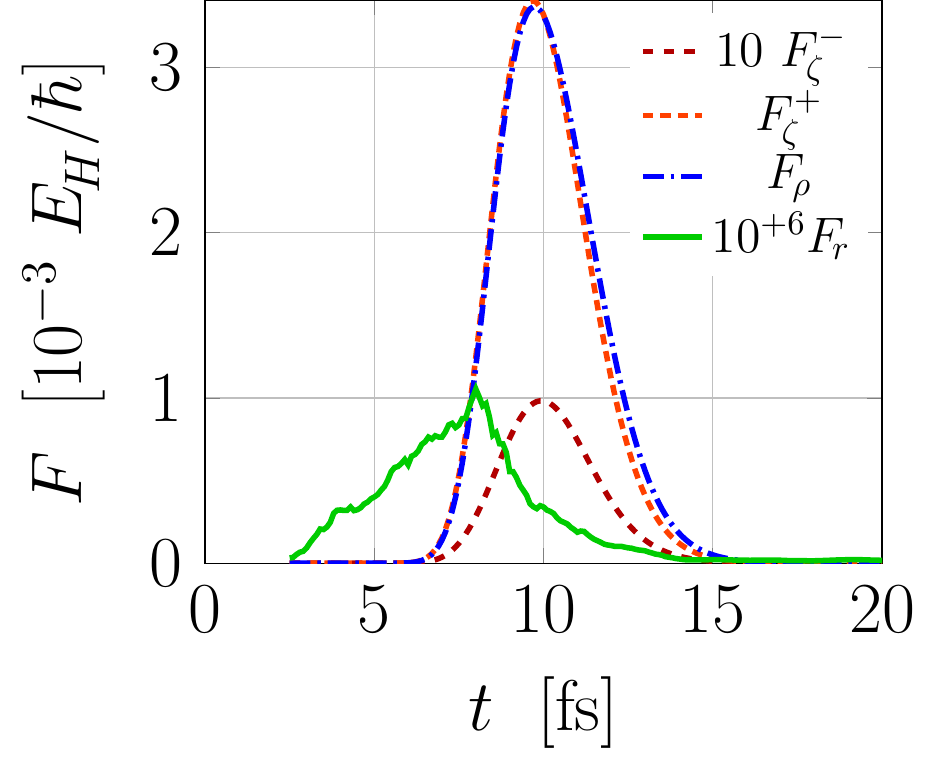}} 
    node[anchor=south west] at (int.north west) {a)}
    node[anchor=south] at (int.north) {Interaction}
    node[anchor=west] (ref) at (int.east)
    {\includegraphics[width=0.33\textwidth]{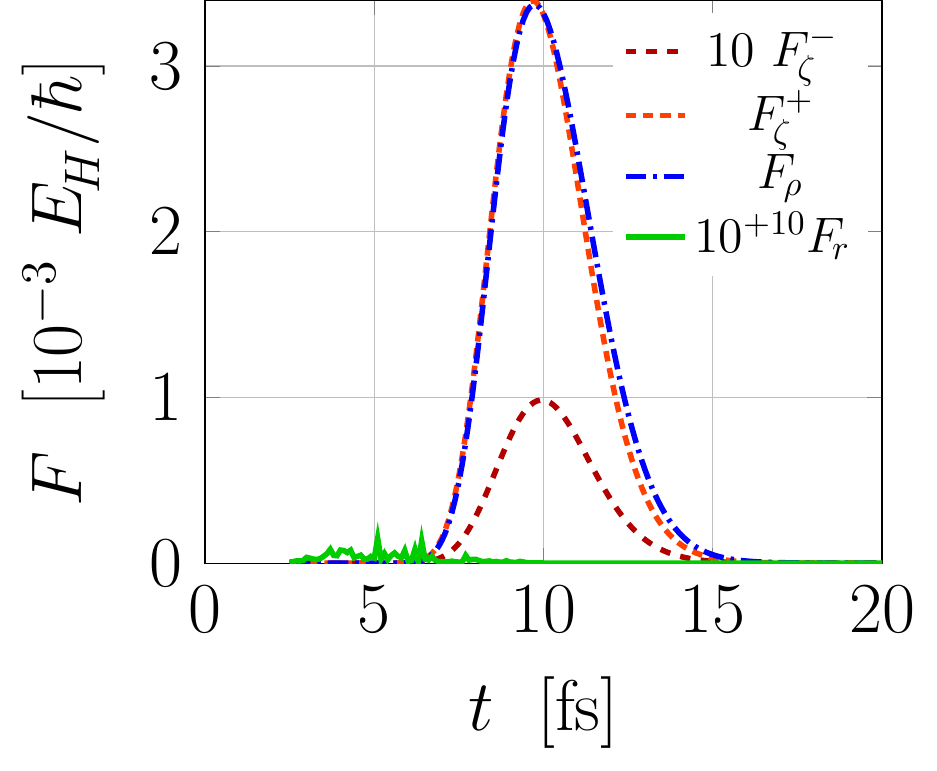}} 
    node[anchor=south west] at (ref.north west) {b)}
	node[anchor=south] at (ref.north) {No Interaction}
    node[anchor=west] (dif) at (ref.east)
    {\includegraphics[width=0.37\textwidth]{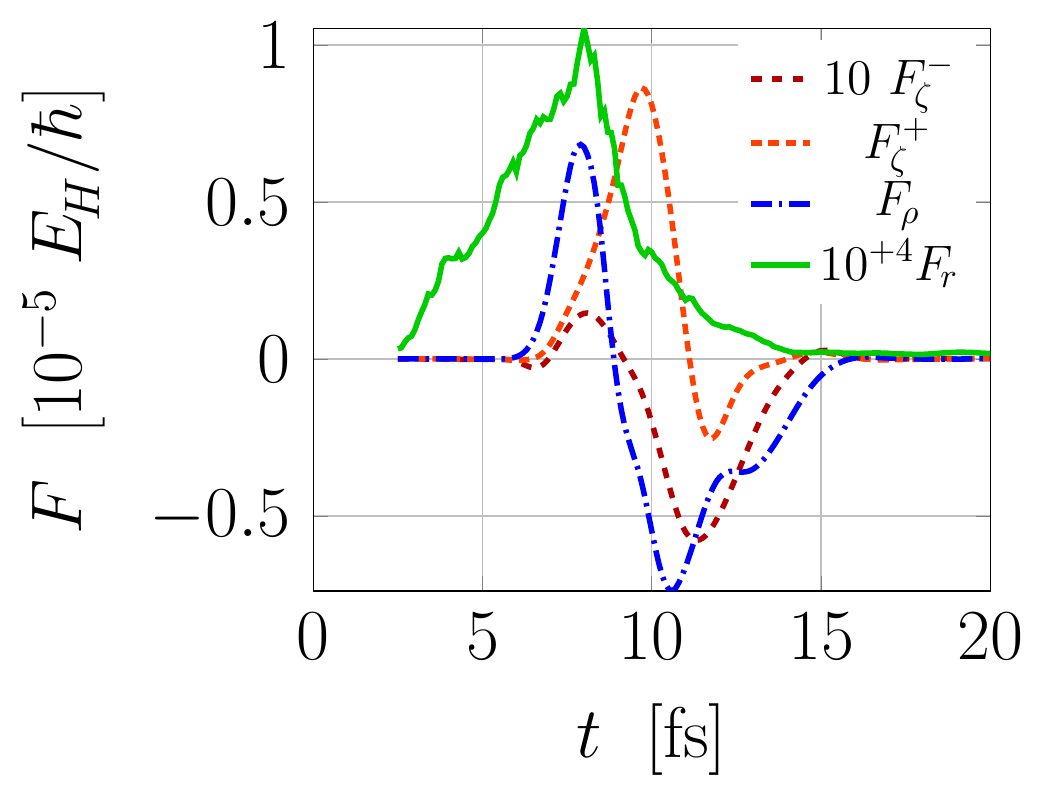}} 
    node[anchor=south west] at (dif.north west) {c)}
	node[anchor=south] at (dif.north) {Difference};
	\hspace*{-.1\textwidth}
\caption{Time-resolved fluxes $F(t)$ of electron probability density reaching the complex absorbing potentials located on the Ba$^{2+}$ site at the longitudinal positions $\zeta=\pm100~\bohr$ (dark-red dashed and orange-red densely dashed) and transverse position $\rho=100~\bohr$ (blue dash-dotted) or respectively from the Rb atom located at $r=100~\bohr$ (green solid). The case for a) fully interacting electrons and b) non-interacting electrons are compared by the difference density c).
	}\label{f:wtt}
\end{figure}

To get more quantitative findings we calculate the electron probability flux $F$ into the different \textsc{cap}s as a function of time $t$. This reveals what fraction of the impinging electron density is scattered for barium and what is in turn emitted from rubidium due to \textsc{icec}. \autoref{f:wtt} shows $F(t)$ at the \textsc{cap} at $\zeta =-100~\bohr$ (dark red dashed), $\zeta =+100~\bohr$ (orange-red densely dashed), $\rho= 100~\bohr$ (blue dash-dotted) and $r=100~\bohr$ (solid green) for interaction turned on in plot a), turned off in plot b), and the difference between the two simulations in plot c). As already seen in \autoref{f:f1} and \autoref{f:f2} (cf.~\autoref*{a:summary},\autoref{f:summary} for a presentation of all graphs together), the scattered electron probability density begins to reach the \textsc{cap} at the barium site after about $6~\fs$. The flux peaks at about $10~\fs$, and after $15~\fs$ almost all probability density is either captured by Ba$^{2+}$ or the \textsc{cap}s. We can also see that the flux in the positive $\zeta$ and in the $\rho$ directions is more than one order of magnitude larger than the backwards direction. This shows that the electron surpasses Ba$^{2+}$ in most cases.

The peak of flux at the rubidium \textsc{cap} is five orders of magnitude less intense which reflects the small probability for \textsc{icec} compared to non-reactive scattering as already observed in the previous section. Furthermore, electron density reaches the Rb~\textsc{cap} much earlier at $t\approx 3~\fs$. The reason for this is that the emitted electron from Rb can have much higher energies than the incoming electron. Moreover, the curve is not smooth, but rather jagged which \txtrev{could possibly indicate} that \textsc{icec} proceeds through a variety of channels, \txtrev{however, at this point we are not aware of how to mathematically distinguish these non-analiticities.}

The right plot of \autoref{f:wtt} shows the difference in the probability flux through the \textsc{cap} surfaces between the simulation with interatomic interaction and with isolated atoms, respectively. We see that the interatomic interaction also influences the scattering at barium. The fluxes are increased at earlier times and reduced at later times during the diffraction event. Integrating over time, the flux at $\zeta = -100~\bohr$ is reduced such that fewer electrons are back\-scattered when the atoms are in interaction. The total flux in the positive $\zeta$ direction, however, is increased, although the overall flux is naturally reduced since some electron probability density is bound at the barium core thanks to the interaction with the rubidium site. The main reduction of the overall diffractive flux at barium arises from the electron probability density registered at the \textsc{cap} surface along $\rho = 100~\bohr$.

\subsection{Reaction probabilities} \label{s:reaction-prob}
We use the presented results to give an estimate for the reaction probability of environment-assisted electron capture. We are looking at time $T= 35~\mathrm{fs}$, which is long after the main scattering event, and there are three possibilities:\\
The first and most common one is that the wave function gets annihilated by the barium \textsc{cap}. This accounts for scattering at the barium but also for delayed scattering -- the electron gets decelerated or temporarily captured at barium but gets reemitted after it takes back the energy from the rubidium site. \\
In the second case the wave function gets annihilated by the \textsc{cap} at the rubidium site; the electron capture at barium leads to the emission of the rubidium electron which gets detected at the rubidium \textsc{cap}. The time integral over the flux at the rubidium \textsc{cap} will then give us the probability for \textsc{icec}.\\
In the last case none of the two electrons has reached their respective \textsc{cap}. This corresponds to the remaining electron density in the system. This can appear either due to scattering of very slow electrons at barium or by electron capture accompanied by the excitation of the rubidium electron. To get the probability for the latter (the first step of \textsc{{\footnotesize2}cdr}) we calculate the remaining electron density in the system at time $T$ but subtract by the case when the interaction is off in order not to count electron scattering of very slow electrons.\\
The time integrated flux into the rubidium \textsc{cap} as presented in \autoref{f:wtt} is
\begin{align}
    \int_{t=0}^T F_{r}^{\text{on}}(t) - F_r^{\text{off}}\ dt = 1.9 \times 10^{-7}. 
\end{align} 
The probability for environment-ionising assisted electron capture, \textit{i.e.} \textsc{icec} as described by Eqs.~\eqref{eq:Ba+Rb+} and \eqref{eq:Ba+*Rb+} is thus $1.9 \times 10^{-5} \%$.
The absolute probability of two electrons remaining in the simulated diatomic system by the time $T$ as presented individually across Figures \autoref{f:f1} -- \autoref{f:f4} adds up to 
\begin{multline}
      \int |\Psi_{\text{on}}(T,\zeta)|^2-|\Psi_{\text{off}}(T,\zeta)|^2 d\zeta 
    = \int |\Psi_{\text{on}}(T,\rho)|^2 -|\Psi_{\text{off}}(T,\rho)|^2 d\rho  \\ 
    = \int |\Psi_{\text{on}}(T,r)|^2-|\Psi_{\text{off}}(T,r)|^2 dr 
    = \left\| \Psi_{\text{on}}(T) \right\|^2 - \left\| \Psi_{\text{off}}(T) \right\|^2 
    = 8.2 \times 10^{-6}.
\end{multline} 
The probability for rubidium-excitation from assisted electron capture as described in context of Eq.~\eqref{eq:Ba+*Rb*} is thus $82 \times 10^{-5} \%$, an order of magnitude larger than the portion that underwent \textsc{icec}.
As seen in the prior sections, electron diffraction dominates the overall simulation with over $99.99\%$ probability and electron capture happens quite rarely.
Similarly small outcomes for assisted electron capture had been observed for the first successful electron dynamics simulations for nanowire-embedded quantum dots.\cite{pont-physrev-2013} In that system it has been established that parameter optimisation can however change the assisted capture contributions by orders of magnitude.\autocite{pont2019-prepare} 
Additionally, the interatomic distance $\distance = 50~\bohr$ is relatively large on an atomic scale. As dipole-dipole mediated energy transfer scales with $\distance^{-6}$,\autocite{gokhberg-jphysb-2009,voitkiv-physrev-2010} this suppresses the probability for environment-assisted capture but is of interest with respect to experimental conditions in ultracold atom systems. Interestingly, the probability for electron capture accompanied by the excitation of the rubidium electron is more than one order of magnitude larger than for capture with emission of the rubidium electron. Our calculation thus predicts that the recombination of rubidium is dominated by the first step of \textsc{{\footnotesize2}cdr} according to Eq.~\eqref{eq:Ba+*Rb*}, rather than the more commonly discussed \textsc{icec} from the Eqs.\eqref{eq:Ba+Rb+} and \eqref{eq:Ba+*Rb+}. 

\subsection{Discussion}\label{s:Discussion}
This paper presents a proof of principle for simulations of dilute diatomic time-resolved environment-assisted electron capture. The physical parameters are not yet optimised to reach high reaction probabilities as was proven possible in the model's predecessor that established simulations of \textsc{icec} dynamics in nanowire-embedded quantum dots.\autocite{pont-physrev-2013,pont2019-prepare} We are not yet aiming to calculate experimental quantities to highest precision at this point but are testing the model in its current stage for viability to simulate electron capture dynamics even at larger experimentally relevant interatomic distances. 
We see that at large distances between the electron and the binding core the current description in dipole-dipole approximation comes close to its validity limit. Therefore, our simulation could overestimate the interaction strength between the atomic sites when the incoming electron is still far away from the barium cation. Additionally for physically more accurate quantities, the binding potential and optimal initial beam parameters will need to be explored.
Our simulation relies currently on a large-core pseudopotential where each atom (ion) carries at most one interacting outer valence electron. 
As our model Hamiltonian reproduces the energy levels (of the $s$ orbitals) very well, the use of the pseudopotential should not falsify the energetics of the process much. Furthermore, our current model using \textsc{mctdh} cannot directly include reaction pathways that include radiative relaxation which in an experiment would be able to compete by itself or support considered reaction channels. \\
The presented research on the electron dynamics of environment-assisted electron capture advances the field in various directions. First of all the application of \textsc{mctdh} with a large \textsc{dvr} grid for the description of continuum electrons was widened from one\autocite{pont-physrev-2013,bande-epjconf-2015,pont-jphys-2016,pont2019-prepare,molle-jcp-2019} (or two in ICD\autocite{Haller2019}) to three Cartesian dimensions. This was possible through recasting the coordinates into a hybrid coordinate system of cylindrical and spherical coordinates for the electron on barium and rubidium, respectively. The second novelty is the electron dynamics being executed in an atomic system, introducing \textsc{icec} in dilute ultracold atom systems. Previously, other atomic systems were only considered non time-resolved at shorter interatomic distances.
When comparing with former electron dynamics results obtained for quantum dots,\autocite{pont-physrev-2013,bande-epjconf-2015,pont-jphys-2016,molle-jcp-2019,pont2019-prepare}
we see that our atomic model allows for significantly more single-electron bound states. Earlier it was already found that more states allow for more \textsc{icec} channels including resonance-enhanced \textsc{icec} pathways.\autocite{pont-physrev-2013,bande-epjconf-2015,pont-jphys-2016,molle-jcp-2019,pont2019-prepare} 

On the one hand, future studies should investigate how the probabilities here are changed if radiative relaxation is included in the simulation. As the interatomic distance is relatively large and the interatomic interaction therefore small, we expect that one center recombination with photon emission could have a non-negligible contribution. Furthermore, the excited barium cation and the rubidium atom could relax by photon emission completing the \textsc{{\footnotesize2}cdr} process. 
On the other hand, none of the parameters were optimised for reaction probabilities at this point of the model development. It has been shown from electron dynamics of \textsc{icec} in nanowire-embedded quantum dots that this can affect reaction probabilities by orders of magnitude.\autocite{pont2019-prepare,pont-physrev-2013} 
\txtrev{For nanowires, initial \textsc{icec} model developmental studies have also returned very small probabilities on the order of $10^{-3}$, however, over time, numerical parameters were tuned in simulations to arrive at significantly higher probabilities of up to $65\%$. 
We expect that the probability of \textsc{icec} in our system therefore can also be majorly optimised given a thorough numerical study of simulation parameters, which will be elaborated on in a future work.
As for experimental observation with the current parameters and probabilities, our current $10^{-5}$ probability has been calculated assuming one reaction partner only. However, typically, one has $10^6$ reaction partners in an atomic cloud. The $10^{-5}$ likelihood per reaction partner could result in a significant collective effect over $10^6$ reaction partners as probabilities add up.
}

One declared goal of the presented research is to pave the way \txtrev{towards} an experimental proof of \textsc{icec} \txtrev{and raise interest in the relevant experimental community}.
Therefore we would like to point out observables that can be straightforwardly compared after improving capture rates through initial parameters as the \textsc{mctdh} algorithm allows to extract considerably more quantities out of the simulation than presented here. For example, the flux at the \textsc{cap}s cannot only be given as a function of time but also as a function of the energy of the annhilated electron density. Another quantity is the transient occupation of electronic levels of the two ions/atoms. Experimentally this is accessible through pump-probe techniques,\autocite{mcpherson1987-595,corkum2007-381,zewail2009-2219} which can be executed in the given ion trap.\autocite{kruekow-physrevA2016,wolf-science2017,sikorsky-natcomm2018,bahrami-physstatsol2019} Beyond this a magnetic field can direct the electrons towards a detector. Here some joint experimental and theoretical research effort will have to go in to the distinction of incoming electrons that were simply scattered and outgoing electrons that are indicators for a successful \textsc{icec}. The distinction may be complicated by the fact that there are numerous available reaction channels. Electron detection was already successfully used to observe the dynamics of ICD in neon dimers\autocite{schnorr2013-093402,schnorr2015-245,takanashi2019-2186} and in iodomethane\autocite{fukuzawa2019-2186} using streaking techniques.\autocite{itatani2002-173903,kienberger2004-817,fruehling2009-523}
Alternatively, the current charge state of barium can already be probed by fluorescence.\autocite{kruekow-physrevA2016}

In order to enhance the comparability of the calculated data with a potential experiment in ultracold atoms beyond a pair of optical tweezers, a many-body extension of the presented model is of interest. In typical atom-ion traps, a single ion is placed in a cloud of atoms.\autocite{bahrami-physstatsol2019,sikorsky-natcomm2018,kruekow-physrevA2016} As here we currently consider only one assisting atom, an extension is planned to account for a many-body environment.\autocite{molle2019-dissertation}

\section{Conclusion} \label{s:Conclusion}
It is known for more than a decade now that electrons can attach to real and artificial atoms or molecules by Coulomb-interaction-mediated energy transfer to the electrons of ionisable (or excitable) species in the environment.\autocite{bande-jphysb2023,gokhberg-jphysb-2009} The dynamics of the electrons engaging in this process coined \textsc{icec} (or \textsc{{\footnotesize2}cdr}) were calculated in this paper for the first time in a truly three-dimensional time-resolved description. The underlying two-electron model system is kept as simple and intuitive as possible to allow for extensive predictions of the process thereby stressing its generalisability. Nonetheless the parametrisation is adjusted for illustration to the dilute diatomic system of an ultracold barium~(II) cation and an assisting rubidium atom at $50~\bohr$ separation. The purpose behind this choice is the experimental feasibility of an ultracold \textsc{icec} experiment, which we seek to promote. 

As the present study considers more energy levels than preceding electron dynamics investigations for other applications, there are numerous additional reaction channels available. We were able to distinguish between electron capture mediated by electron emission and capture mediated by electron excitation at the assisting atom, and we showed that assisting excitation dominates over emission in the presented system. We thereby found that \textsc{mctdh} is a viable tool to simulate the time-resolved multidimensional electron dynamics of environment-assisted electron capture processes even at larger experimentally relevant distances.

\appendix
\section*{Author Contributions}
The current version of the text in this manuscript has been written by AM, JPD and VN with contributions by AB.
Prior versions of the text were written by AM with contributions from AB who also acted as a supervisor.

The theoretical derivation of the presented model was undertaken by AM with Oriol Vendrell as an advisor during the first exploratory steps in the model development.

AM composed the model description for the electron-dynamical simulations presented here and conducted the simulations. Exploratory work on robust computational parameters for these simulations was undertaken by VN and NK under supervision by AM embedded in the research team of AB.
The procurement of simulation data, its analysis and graphical representations in figures were undertaken by AM. JPD contributed to the analysis of the data, as well as to their final graphical representation together with VN. 

Financial support for this project was acquired by AB throughout model development and completion of simulation and prior manuscript versions. Funding for the revision of this manuscript was acquired by AM.
Additionally, AB and Oriol Vendrell acted in editorial capacity throughout this project.

\section*{Acknowledgements}
AM and JPD are grateful for financial support through the junior postdoctoral fellowship mandate 1232922N by the Research Foundation -- Flanders (\textsc{fwo}).
AM, VN, and AB thank the Volkswagen Foundation for the primary funding of this project through the Freigeist grant no.~89525 (grant holder AB).
The project benefited from the three-month research stay of AM hosted by Oriol Vendrell at Aarhus University in Winter 2017/18, which was financially supported through the \textit{PhD Student Research abroad program} of the Helmholtz-Zentrum Berlin. 
The internship of NK was supported by the German Academic Exchange Service (\textsc{daad}) through the \textsc{rise} program.
The internship of VN was partially supported by an internship bursary within the Moritz-Heyman Scholarship by the University of Oxford.

AM and AB gratefully acknowledge their respective invitations as speakers at the launch meeting of the \textsc{mctdh} International Research Network (\textsc{irn} \textsc{mctdh}) held at \textsc{ismo}, Paris-Saclay in September 2021 and the financial support of their respective contributions to that meeting by the Quantum Dynamics Network which enabled a fruitful in-person exchange on the manuscript.
Moreover, AM expresses their gratitude for the Fulbright Schuman-Program Award 2022/23 issued by the Commission for Educational Exchange between the United States, Belgium and Luxembourg. This recognition does not imply that the Government of the United States or any agency representing it has endorsed the conclusions or approved the contents of this publication.

\printbibliography

\providecommand{\xv}[1]{\ensuremath {\bigl< \hat #1 \bigr>} }
\providecommand{\bra}[1]{\ensuremath {\bigl< #1 \bigr|} }
\providecommand{\ket}[1]{\ensuremath {\bigl| #1 \bigr>} }
\providecommand{\braket}[2]{\ensuremath{\bigl< #1 \bigm| #2 \bigr>}}
\providecommand{\ketbra}[2]{\ensuremath {\bigl|{ #1 }\bigm>\!\bigm<{ #2 }\bigr|} }
\providecommand{\op}[1]{\ketbra{#1}{#1}}
\providecommand{\mel}[3]{ \ensuremath{\bigl< #1 \bigm| #2 \bigm| #3 \bigr>} }

\providecommand{\1}{ {\mathbb{1}} }
\providecommand{\N}{\mathcal{N}}
\providecommand{\Proj}{ {\hat{\mathbb{P}}} }
\providecommand{\dd}[2][]{{d}^{#1\!}{#2}~}
\providecommand{\jacobian}{\mathcal{J}}

\section{Derivation of Jacobian-Normalised Operators}
\label{a:jacobian}
This appendix outlines the mathematical derivation of the employed kinetic energy operators which have been normalised with respect to the individual Jacobian of the chosen coordinate subsystems for barium and rubidium.
The concept of a (Jacobian-)normalised kinetic energy operator arises naturally though implicitly from the mathematical solution for stationary states of the hydrogen atom from the Schrödinger equation in spherical polar coordinates.\autocite{landaulifshitz, feynman3, schrodinger} It implies a simultaneous normalisation of the wavefunction's representation and the kinetic operators for curvilinear coordinates and is sometimes called a normalisation within the \textit{Dirac convention}.\autocite{nauts-mp1985}

We will first motivate here the generalised concept of such normalisation with respect to the Jacobian of a chosen coordinate system. This was previously shown to justify the model presented here.\autocite{molle2019-dissertation}
We will then illustrate this normalisation procedure for the more commonly found spherical polar coordinate system as employed for the rubidium subsystem (cf. Eq.~\eqref{eq:TRb}), before we illustrate the derivation of the Jacobian-normalised kinetic energy operator for a cylindrical coordinate system, as used for the barium cation in the presented model (cf. Eq.~\eqref{eq:TBa}).
Throughout this appendix we confine ourselves to a single particle, a single electron in particular, and its quantum mechanical description in the three-dimensional physical space.

\paragraph{}
The same point $\vec r$ in three-dimensional space can be described by different vectors based on the choice of coordinate systems. The choice of coordinate system will have an impact on the description of partial differential equations and integrals that form the foundation of the quantum mechanics employed in this work.

Let the conventional Cartesian coordinate system be represented by $\vec r = (x,y,z)$ and as an alternative choice, potentially non-Cartesian, be given by $\vec r = (x_1,x_2,x_3)$. The Jacobian determinant, denoted here as $|\jacobian|^2$, relates the two coordinate systems through their respective partial derivatives. It is thus given by\autocite{liu-jma2022,magnus2019}
\begin{equation}\label{eq:jacobian}
|\jacobian|^2:=\det\left(\frac{\partial(x,y,z)}{\partial(x_1,x_2,x_3)}\right)
.
\end{equation}
Furthermore, the differential volume element $\dd[3]{\vec r}$ is essential in multivariate integration. Its description depends on the chosen coordinate system. The Jacobian determinant arises as a correcting factor within the integral when changing coordinate systems (e.g. from Cartesian to non-Cartesian). Formally, it is given in both coordinate systems as
\begin{equation}\label{eq:dV}
\dd[3]{\vec{r}}=\dd{x} \dd{y} \dd{z} = \dd{x_1} \dd{x_2} \dd{x_3} \; |\jacobian|^2 \;.
\end{equation}
The Jacobian determinant is thereby a scalar function of the underlying three-dimensional vector space. This is similar in its nature to the wave function $\psi$ at any instant in time.

\paragraph{}
Within the Copenhagen interpretation of quantum mechanics, we consider the inner product of a wave function $\psi$ with itself at any instant in time as representation of a probability. Furthermore, assuming that a particle is present somewhere in space and is represented by this wave function $\psi$, we request that this wave function is normalised as
\begin{equation}
1 \equiv \iiint\dd[3]{\vec{r}} ~\psi^*\psi,
\end{equation}
where $\psi*$ is the complex conjugate of the wave function.
In other words, the probability must reflect our assumption that the particle is somewhere in space at any time.
This has to hold for any choice of coordinate system. Changing the description must not affect the underlying physics:
\begin{equation}
\iiint\dd[3]{\vec{r}} ~\psi^*\psi = \iiint\dd x  \dd y  \dd z   ~|\psi|^2 = \iiint\dd{x_1} \dd{x_2} \dd{x_3} ~| \jacobian|^2 \; |\psi|^2 \;.
\end{equation}

\paragraph{}
Numerical integration algorithms are insensitive to our physical interpretation of the chosen coordinate system. That means, the algorithm does not incorporate the Jacobian determinant automatically for an arbitrary choice of coordinates. This necessitates that $|J|^2$ is included within an adapted wave function $u_{\!\jacobian}$ such that
\begin{equation}
\iiint \dd{x} \dd{y} \dd{z} ~|\psi(x,y,z)|^2 
=: \iiint\dd{x_1} \dd{x_2} \dd{x_3} 
	 ~|u_{\!\jacobian}(x_1,x_2,x_3)|^2 \;.
\end{equation}
The Jacobian determinant and the wave function are both scalar functions on the complex plane. This implies that they commute with each other and we can identify the adapted wave function as the product of functions
\begin{equation}
u_{\!\jacobian}(x_1,x_2,x_3) = \jacobian(x_1,x_2,x_3) \; \psi(x_1,x_2,x_3)
\;. \label{eq:uJ}
\end{equation}
By this, we incorporate the non-Cartesian aspect of the volume element within the wave function. It can therefore be considered \textit{Jacobian-normalised}. In the fundamental example of the radial wave function $R(r)$ for the hydrogen atom, textbook literature conventionally speaks of the \textit{normalised wave function} $u(r)=r \;R(r)$.\autocite{landaulifshitz,feynman3} While this is ambiguous in its choice of wording addressing (Jacobian-)normalisation, it is commonly used.

 \paragraph{}
 Beyond the interpretation of a probability density contained within the square modulus $|\psi|^2$ of the wave function, quantum mechanics relies on expectation values of operators. In order to allow for a physical interpretation of those, we require them to be equally unchanged by a transformation of coordinates.
 Incorporation of the Jacobian determinant into our adapted wave function $u_{\!\jacobian}$ will force us to also adapt our description of such an operator.
 
 Let $\hat O$ denote an operator acting to the right on a wave function as $(\hat O \psi)$. An example would be a differential operator.
 Its expectation value at a given instant in time is then
 given by
 \begin{equation}
 \xv{ O }
 =\iiint \dd[3]{ \vec r} ~\psi^* ~(\hat{O}\psi) 
 = \iiint\dd{x} \dd{y} \dd{z} ~\psi^* ~(\hat{O}\psi)
 = \iiint\dd{x_1} \dd{x_2} \dd{x_3} |\jacobian|^2  ~\psi^* ~(\hat{O}\psi)
 \;.
 \end{equation}
 Since we numerically require to work with the (Jacobian-)normalised wave function $u_{\!\jacobian}$, we seek an associated adapted operator $\hat O_{\!\jacobian}$, such that
 \begin{equation}
 	 \iiint\dd{x_1} \dd{x_2} \dd{x_3} ~{u_{\!\jacobian}}^* ~( \hat O_{\!\jacobian} ~u_{\!\jacobian} )
 	 :=
 	 \iiint\dd{x_1} \dd{x_2} \dd{x_3} |\jacobian|^2  ~\psi^* ~(\hat{O}\psi)
 	 \;.
 \end{equation}
 We identify the integrands and commute the scalar functions $J$ and $\psi^*$ on the right hand side of the equation,
 \begin{equation}
 	(\jacobian \psi)^* ~( \hat O_{\!\jacobian} ~(\jacobian \psi) ) 
 	=
 	{u_{\!\jacobian}}^* ~( \hat O_{\!\jacobian} ~u_{\!\jacobian} ) 
 	=
 	|\jacobian|^2  ~\psi^* ~(\hat{O}\psi)
 	=
 	(\jacobian \psi)^*  ~\jacobian (\hat{O}\psi)
 	\;,
 \end{equation}
 which leaves us with a typical operator transformation relation
 \begin{equation}
	 \hat O_{\!\jacobian} ~(\jacobian \psi) 
	 =
	 \jacobian (\hat{O}\psi)
	 \;.
 \end{equation}
 Under the assumption that an inverse function $\jacobian^{-1}$ exists, such that
 \begin{equation}
	 \jacobian ~\jacobian^{-1} \equiv 1
	 \;,
 \end{equation}
 we multiply it from the right to reach
 \begin{equation}
 \hat O_{\!\jacobian} ~(\jacobian \psi) \jacobian^{-1}
 =
 \jacobian (\hat{O}\psi) \jacobian^{-1}
 \;.
 \end{equation}
 The wave function $\psi$ and the function $\jacobian^{-1}$ commute because they are both scalar functions.
We thereby identify the  (\textit{Jacobian-})\textit{normalised operator} formally as
 \begin{equation}
 \hat O_{\!\jacobian} 
 =
 \jacobian \hat{O} \jacobian^{-1}
 \;. \label{eq:OJ}
 \end{equation}

 \paragraph{}
 The operator governing the evolution of a given wave function with time is the Hamiltonian operator itself through the time-dependent Schr\"odinger equation. We restrict ourselves to the description of an atom as employed in the presented model where a single electron can be described in its own three-dimensional subspace with an attractive pseudopotential $\hat V$ at its origin. The electronic Hamiltonian operator $\hat h$ is composed of the kinetic energy operator $\hat T$ of the electron and the potential energy operator $V$, such that
 \begin{equation}
 	\hat h = \hat T + V \;.\label{eq:hTV}
 \end{equation}
 The potential energy operator is a scalar function in three-dimensional space, unlike the kinetic energy operator. Therefore it commutes with the wave function and the Jacobian determinant and function $\jacobian$.
 The electron dynamics thus arises from the time-dependent Schr\"odinger equation,
 \begin{equation}
 	 i \hbar \tfrac\partial{\partial t} \psi = \hat h \psi
 	\;.
 \end{equation}
 The generalised nature of numerical integration and differentiation algorithms demand, however, that the wave function and the Hamiltonian are expressed in a Jacobian-normalised form to correct for non-Cartesian coordinate choices.
 We are therefore required to formulate our model in terms of $u_{\!\jacobian}$ and $\hat h_{\!\jacobian}$, to allow for a numerical solution of
 \begin{equation}
 	 i \hbar \tfrac\partial{\partial t} u_{\!\jacobian} = \hat h_{\!\jacobian} ~u_{\!\jacobian}
 	\;.
 \end{equation}
 The Jacobian-normalised wave function $u_{\!\jacobian}$ can be straightforwardly determined from Eq.~\eqref{eq:uJ} if $\psi$ is known for an initial instant in time.
 The scalar nature of the potential energy operator $V$ and its commutation with the function $J$ results in the identity
 \begin{equation}
 	V_{\!\jacobian} = \jacobian V \jacobian^{-1}
 		= \jacobian \jacobian^{-1} V = V.
 \end{equation} 
 The Jacobian-normalised Hamiltonian $\hat h_{\!\jacobian}$ implies from Eq.~\eqref{eq:OJ} and Eq.~\eqref{eq:hTV} thereby a necessity for a normalised kinetic energy operator $\hat T_{\!\jacobian}$ as
 \begin{equation}
 	h_{\!\jacobian} = \hat T_{\!\jacobian} + V \;.
 \end{equation}
 
 For a single particle, the kinetic energy operator is commonly given in terms of the second spatial differential as
 \begin{equation}
 	\hat T \psi = \left(\tfrac {-\hbar^2\nabla^2}{2m} \right) \psi,
 \end{equation}
 where $m$ is the particle's mass, here the electron mass, and $\nabla^2$ is to be expressed in the chosen coordinate system.
 From here, we arrive at the question for the Jacobian-normalised kinetic energy operator $T_{\!\jacobian}$, particularly in spherical and cylindrical coordinates, as necessitated by the developed model,
 \begin{equation}
	 \hat T_{\!\jacobian} ~u_{\!\jacobian}
	 = \jacobian \hat T \jacobian^{-1} ~u_{\!\jacobian}
	 = -\tfrac{\hbar^2}{2m} \jacobian \nabla^2 ~(\jacobian^{-1} u_{\!\jacobian})
	 \;.
 \end{equation}
 
\subsection{Normalised Spherical Polar Kinetic Energy Operator}\label{a:spherical}

Spherical polar coordinates are a natural choice for atomic systems. The attractive potential energy of an electron is spherically symmetric around a nucleus. For the hydrogen atom as a pedagogical example, the transition to spherical coordinates allows for the separation of variables into a radial and an angular partial differential equation and the solution for stationary states in those lower-dimensional problems. As we employ a pseudopotential for our model as an effective atomic binding potential to the outer valence electron, we arrive at a hydrogen-like description for the rubidium atom's outer modelled valence electron.

The spatial vector in spherical polar coordinates is given by $\vec r = (r,\theta,\varphi)$ as employed for the rubidium atom in the presented quantum dynamical model. The Cartesian coordinates can be expressed as functions of the spherical polar coordinates as
\begin{align}
	x(\vec r) &= r \sin \theta \cos\varphi ,\\ 
	y(\vec r) &= r \sin \theta \sin\varphi ,\\
	z(\vec r) &= r \cos \theta \;,
\end{align}
which leads to the Jacobian determinant
\begin{equation}
	|\jacobian (r,\theta,\varphi)|^2 = 
	\begin{vsmallmatrix}
		\tfrac{\partial x}{\partial r} & 
			\tfrac{\partial y}{\partial r} & 
			\tfrac{\partial z}{\partial r} \\
		\tfrac{\partial x}{\partial \theta} &
			\tfrac{\partial y}{\partial \theta} &
			\tfrac{\partial z}{\partial \theta} \\
		\tfrac{\partial x}{\partial \varphi} &
			\tfrac{\partial y}{\partial \varphi} &
			\tfrac{\partial z}{\partial \varphi}
	\end{vsmallmatrix}
	= r^2 \sin\theta
	\;.
\end{equation}
We do have a freedom of complex phase in our choice of the scalar function $\jacobian$.

\paragraph{}
However, we point out here, that $\sin\theta\,\dd\theta$ can be formally written as $\dd{(-\cos\theta)}$. The cosine function is a bijective map from the polar angle $\theta$ ranging from $0$ to $\pi$ onto the interval from 1 to -1. That means there is a non-ambiguous coordinate transformation possible from angle $\theta$ to its cosine $\cos\theta$.
We therefore choose a description of the spatial vector $\vec r$ in spherical polar coordinates in terms of 
$\vec r = (r, -\cos\theta, \varphi)$ instead. 
This relates back to the Cartesian coordinates as
\begin{align}
x(\vec r) &= r \sqrt{1-\cos^2\theta} \cos\varphi ,\\ 
y(\vec r) &= r \sqrt{1-\cos^2\theta} \sin\varphi ,\\
z(\vec r) &= r \cos \theta \;,
\end{align}
which leads to a purely radial Jacobian determinant  
\begin{equation}
|\jacobian_s|^2 = 
\frac{ \partial (x,y,z) }{ \partial (r,-\cos\theta,\varphi) }
= r^2
\;.
\end{equation}

While we have a freedom of choice within the complex phase of $\jacobian_r$, it appears sufficient for the moment to choose the simple linear function
\begin{equation}
	\jacobian_s = r
	\;.
\end{equation}
This gives rise to the inverse function
\begin{equation}
	\jacobian_s^{-1} = r^{-1},
\end{equation}
which is well defined except on the singular points at the origin $r=0$. This is of no further concern to us, as any volumetric integration of the singular point alone results in a zero volume.
We note that we have arrived at a $1/r$ dependence which also includes the essential feature of the Coulomb binding potential.

The textbook expression for the Laplacian $\nabla^2 f$ of an arbitrary complex scalar function $f$ is conventionally expressed as \autocite[p.~2]{cohentannoudji}
\begin{equation}
	\nabla^2 f =
		r^{-1} 
			\partial_r^2 (r f)
		+ r^{-2} \left[ 
		(\sin\theta)^{-1} 
			\partial_\theta (\sin\theta \partial_\theta f)
		+ (\sin\theta)^{-2} 
			\partial_\varphi^2 f
			\right]
	\;.
 \label{eq:s-laplacian}
\end{equation}
The angular contributions within the square brackets reflect the squared orbital angular momentum operator\begin{equation}
\hbar^2 \hat \ell ^2  := -\hbar^2 \left[(\sin\theta)^{-1} 
\partial_\theta \sin\theta \partial_\theta
+ (\sin\theta)^{-2} 
\partial_\varphi^2 \right]
\;,
\end{equation}
which can be formally simplified 
to
\begin{equation}
\hat \ell ^2  = - \left( \sin\theta \right)^{-2} \left[
	\partial_{-\cos\theta}^2 + \partial_\varphi^2
\right] 
\;.
\end{equation}
The angular partial derivatives $\partial_{-\cos\theta}$ and $\partial_\varphi$ are independent of the radial coordinate itself and therefore commute with $r$ and $1/r$. In other words, the angular momentum operator $\hat \ell$ remains unaffected by Jacobian-normalisation with $\jacobian_s$. Furthermore, the angular momentum operator is implemented in the MCTDH software for the chosen discrete variable representation for the angular coordinates and can therefore be used in the formulation of the working equation for the kinetic energy operator here.
For spherical polar coordinates, the normalised kinetic energy operator is therefore
\begin{equation}\label{eq:Ts}
\hat{T}_{\!s} ~u_s
= 
\jacobian_s \hat T \jacobian_s^{-1} ~u_s
= 
\tfrac{-\hbar^2}{2m} \;r \nabla^2 r^{-1} ~u_s
=
 \tfrac{-\hbar^2}{2m} \left[ \partial_r^2 - r^{-2} ~\hat \ell^2	
\right] ~u_s
 \;.
\end{equation}

\paragraph{}
This formal normalisation procedure for the kinetic energy operator is commonly used implicitly for the hydrogen eigenvalue problem where the Laplacian is rarely presented in its canonical form of Eq.~\eqref{eq:s-laplacian}.
The radial Schr\"{o}dinger equation for the Coulomb potential after separation of variables is rather already shown in Jacobian-normalisation as\autocite[Eq.~(32.10),(36.2)]{landaulifshitz}
\begin{equation}
\frac{d^2 u_{n,\ell}(r)}{\dd{r}^2} 
+\frac{2m}{\hbar^2}\left(
E_n  + \frac{\kappa}{r} 
- \frac{\ell(\ell+1)}{r^2}
\right)u_{n,\ell}(r)
=0,
\end{equation}
which is using $\hat T_s$, and where $n$ and $l$ are the principal and the angular momentum quantum number, $E_n$ is the associated eigenenergy, and $\kappa$ represents the Coulomb constant in this context.
This is in turn related to the associated Laguerre differential equation\autocites[(36.4)]{landaulifshitz}[(22.6.15)f]{abramowitzstegun}
\begin{equation}
0 = x u''+ (x+1)u' + \left(n + \tfrac{\ell}{2}+1-\frac{\ell^2}{4x}\right)u \,,
\end{equation}
for different boundary conditions imposed by angular momentum quantum number $\ell$. Its solutions of the form\autocites[(22.6.15)f]{abramowitzstegun} \\
\begin{equation}
u =  x^{{\ell}/{2}} \;e^{-x} \; L_n^{\ell}(x)\,\text{, where }\;
L_n^\ell(x) = \sum\limits_{j=0}^{n}  \binom{n+\ell}{n-j} \frac{(-x)^j}{j!}
\end{equation}
 are the associated Laguerre polynomials.\autocites[(22.3)]{abramowitzstegun}
 They provide the essential ingredients to the numerical implementation of the
 Laguerre \textsc{dvr}s in spherical and cylindrical coordinates used in \autoref{s:compdetails}.

\subsection{Normalised Cylindrical Kinetic Energy Operator}\label{a:cylindrical}
For the cylindrical polar coordinate system as employed in the model for the barium subsystem, we use the coordinate vector $\vec r = (\zeta, \rho, \varphi)$, where $\zeta$ represents the cylindrical axis, $\rho$ represents the transverse radial component with respect to $\zeta$ and $\varphi$ is the azimuthal angle on the transverse plane.
We note that for the presented case of simulation, the electron beam direction is perpendicular to the interatomic axis. The interatomic axis is chosen parallel to the local $z$ axis as is conventional. The electron beam axis is thus aligned here with the $y$ axis. This leads to a permutation of labels $(x,y,z)$ with respect to the usual expression of the Cartesian coordinates through the cylindrical ones:  
\begin{align}
	y( \vec r ) &= \zeta, \\
	z( \vec r ) &= \rho \cos\varphi, \\
	x( \vec r ) &= \rho \sin\varphi
	\;.
\end{align}
We arrive at the usual Jacobian determinant for cylindrical polar coordinates
\begin{equation}
 |\jacobian_c|^2 = \rho \,,
\end{equation}
such that we choose the functions
\begin{align}
	\jacobian_c^{\pm 1} = \rho^{\pm\frac12} \,.
\end{align}
Moreover, the 
Laplacian of an arbitrary complex scalar function $f$ in cylindrical polar coordinates is commonly expressed as\autocite{cohentannoudji}
\begin{equation}
	\nabla^2 f = 
		\partial_\zeta^2 f
		+
		\rho^{-1}
			\partial_\rho (\rho \partial_\rho f)
		+ \rho^{-2}
			\partial_\varphi^2 f
		\;.
\end{equation}
We note that $\partial_\zeta$ and $\partial_\varphi$ commute with the scalar functions of $\rho$. This includes the functions $\jacobian_c^{\pm1}$. 
Moreover, we identify the angular momentum operator along the cylindrical axis $-i\hbar\partial_{\varphi_{\!\Ba}}=:\hbar\hat{\ell}_\zeta$.
For cylindrical polar coordinates, the normalised kinetic energy operator is therefore
\begin{equation}
\hat{T}_{\!c} ~u_c
= 
\jacobian_c \hat T \jacobian_c^{-1} ~u_c
= 
\tfrac{-\hbar^2}{2m} \;\rho^{\frac12} \nabla^2 \rho^{-\frac12} ~u_c
=
\tfrac{-\hbar^2}{2m} \left[
	\partial_\zeta^2
	+ 
	\rho^{-\frac12}
		\partial_\rho \rho \partial_\rho
		\rho^{-\frac12}
+ \rho^{-2}
		\partial_\varphi^2 
\right] ~u_c
\;.
\end{equation}
We aim to simplify the term with respect to the cylindrical transverse radial coordinate,
\begin{align}
	\rho^{-\frac12}
	\partial_\rho \rho \partial_\rho
	\rho^{-\frac12} u_c
	&=
	\rho^{-\frac12}
	\partial_\rho \left[\rho (\partial_\rho
	\rho^{-\frac12})
	+
	\rho^{+\frac12} \partial_\rho\right]
	 u_c
\\
	 &=
	-\tfrac12 \rho^{-\frac12}
	\partial_\rho  
	(\rho^{-\frac12} u_c)
	+
	\left[
	\rho^{-\frac12}
	(\partial_\rho \rho^{+\frac12}) \partial_\rho
	+
	\partial_\rho^2
	\right] u_c 
\\
	&=
	\left[
	-\tfrac12 \rho^{-\frac12}
		(\partial_\rho  
		\rho^{-\frac12})
	-\tfrac12 \rho^{-1}
		\partial_\rho  
	 +
	 \rho^{-\frac12}
	 (\partial_\rho \rho^{+\frac12}) \partial_\rho
	 +
	 \partial_\rho^2
	 \right] u_c 
\\
	&=
	\left[
	-\tfrac14 \rho^{-2}
	-\tfrac12 \rho^{-1}
	\partial_\rho  
	+\tfrac12\rho^{-1}\partial_\rho
	+
	\partial_\rho^2
	\right] u_c 
=
	\left[
		\partial_\rho^2
		-\tfrac14 \rho^{-2}
	\right] u_c 
\,,
\end{align}
such that we reach the final equation of Jacobian-normalised cylindrical polar kinetic energy operator as used for the presented quantum dynamical model for the barium subsystem,
\begin{equation}\label{eq:Tc}
\hat{T}_{\!c} ~u_c
= 
\tfrac{-\hbar^2}{2m} \left[
\partial_\zeta^2
+ 
\partial_\rho^2
+ \rho^{-2} \left(
\partial_\varphi^2
-\tfrac14 \right) 
\right] ~u_c \,.
\end{equation}
 
\clearpage
\section{Summary of Time-Dependent Simulation Data}\label{a:summary}
\begin{figure*}[h!]\centering
	\addtolength\fboxsep{-2pt}
	\caption{Probability Density Evolution Summary} \label{f:summary}
	\hspace*{-.1\textwidth}\tikz\node (int){\includegraphics[width=0.4\textwidth]{f1intden}} node[anchor=north] at (int.north) {Interaction}node[anchor=west, xshift=-1em] (ref) at (int.east){ \includegraphics[width=0.4\textwidth]{f1refden}
		} node[anchor=north] at (ref.north) {No Interaction}node[anchor=west, xshift=-1em] (dif) at (ref.east){ \includegraphics[width=0.4\textwidth]{f1difden}
	} node[anchor=north] at (dif.north) {Difference} node[anchor=north, yshift=3em] (f2 int) at (int.south){\includegraphics[width=0.4\textwidth]{f2intden} }node[anchor=west, xshift=-1em] (f2 ref) at (f2 int.east){\includegraphics[width=0.4\textwidth]{f2refden}}node[anchor=west, xshift=-1em] (f2 dif) at (f2 ref.east){\includegraphics[width=0.4\textwidth]{f2difden}}node[anchor=north, yshift=3em] (f4 int) at (f2 int.south){\includegraphics[width=0.4\textwidth]{f4intden}}node[anchor=west, xshift=-1em] (f4 ref) at (f4 int.east){\includegraphics[width=0.4\textwidth]{f4refden}}
    node[anchor=west, xshift=-1em] (f4 dif) at (f4 ref.east){\includegraphics[width=0.4\textwidth]{f4difden}}node[anchor=north, yshift=1em, xshift=-.5em] (flx int) at (f4 int.south){\includegraphics[width=0.31\textwidth]{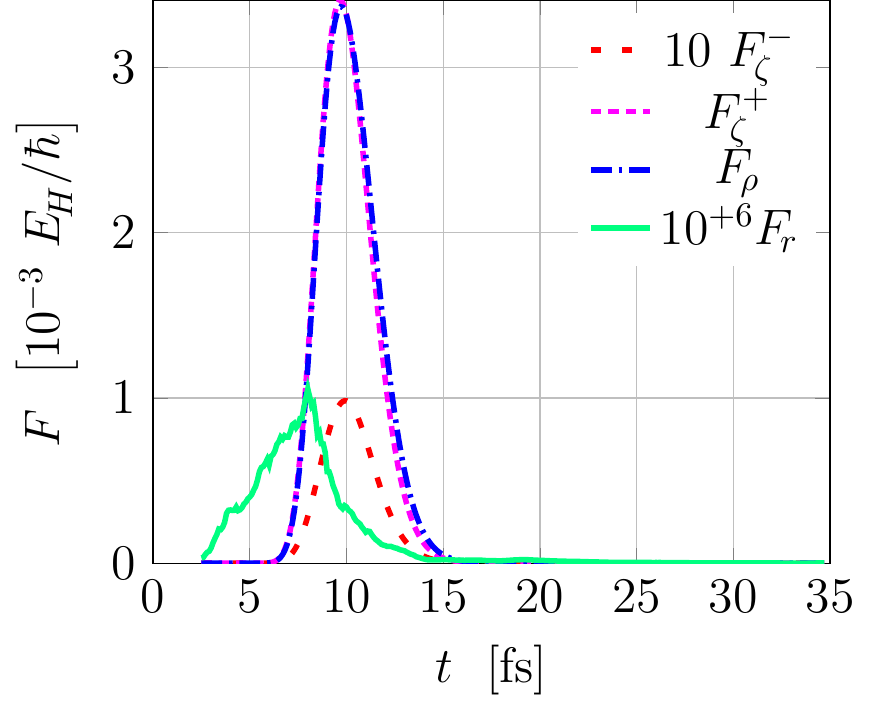}}node[anchor=north, yshift=1em, xshift=-0.5em] (flx ref) at (f4 ref.south){\includegraphics[width=0.31\textwidth]{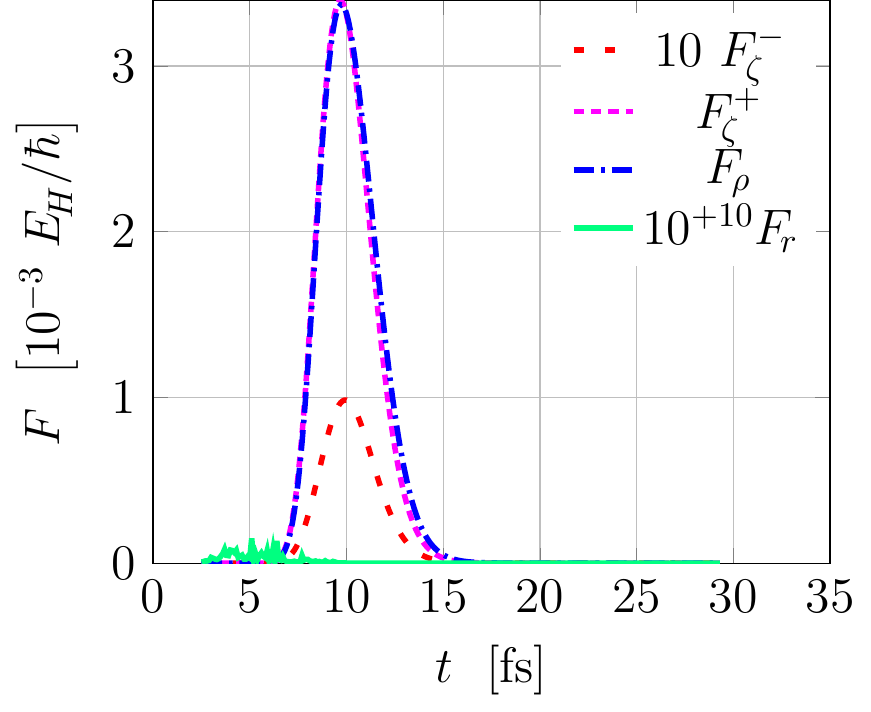}}
node[anchor=north, yshift=0.9em, xshift=-0.75em] (flx dif) at (f4 dif.south){\includegraphics[width=0.32\textwidth]{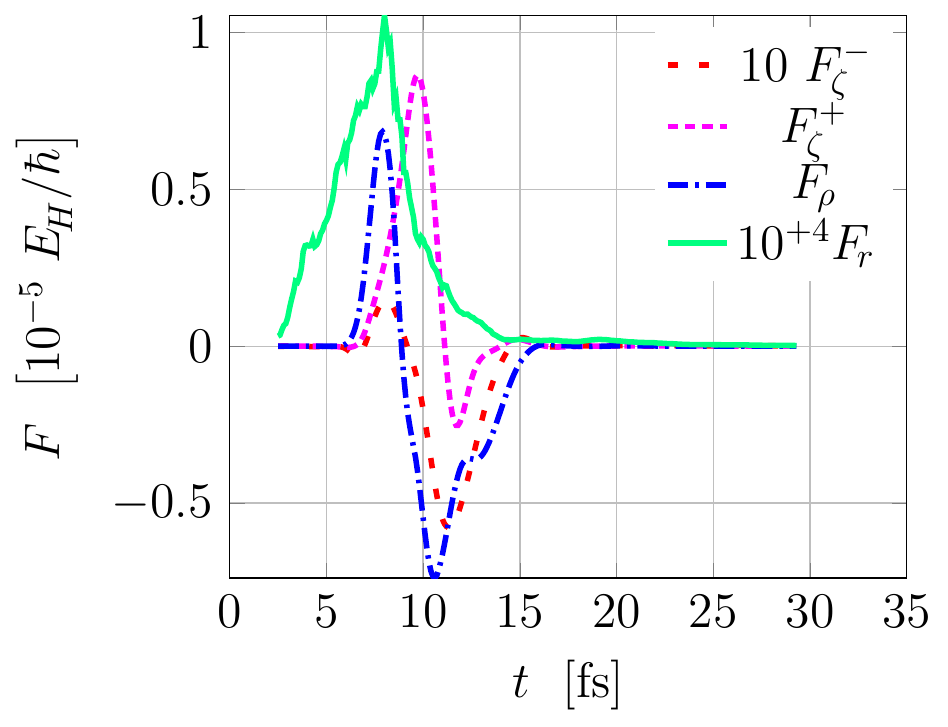}};
\end{figure*}\vspace*{-10em}
\end{document}